\documentclass[%
 reprint,
nofootinbib,
 amsmath,amssymb,
 aps,
]{revtex4-2}

\usepackage{graphicx}
\usepackage{dcolumn}
\usepackage{bm}
\usepackage{subfigure}
\usepackage{float}

\usepackage{tikz}
\usetikzlibrary{decorations}
\usepackage{ragged2e}

\begin{document}

\preprint{APS/123-QED}

\title{Microwave sink using plasma-based localized surface plasmons}

\author{Benjamin Fromont}
\email{benjamin.fromont0911@gmail.com}
\affiliation{ISAE-SUPAERO, Université de Toulouse, France}

\author{Julien De Rosny}
\affiliation{Institut Langevin, ESPCI Paris, Université PSL, CNRS, 75005 Paris, France}

\author{Romain Pascaud}
\affiliation{ISAE-SUPAERO, Université de Toulouse, France}

\author{Nicolas Lebbe}
\affiliation{LAPLACE, Université de Toulouse, CNRS, INPT, UPS, Toulouse, France}

\author{Olivier Pascal}
\affiliation{LAPLACE, Université de Toulouse, CNRS, INPT, UPS, Toulouse, France}

\author{Jérôme Sokoloff}
\affiliation{LAPLACE, Université de Toulouse, CNRS, INPT, UPS, Toulouse, France}

\author{Laurent Liard}
\affiliation{LAPLACE, Université de Toulouse, CNRS, INPT, UPS, Toulouse, France}

\author{Théo Delage}
\affiliation{ENAC, Université de Toulouse, Toulouse, France}

\author{Antoine Saucourt}
\affiliation{ISAE-SUPAERO, Université de Toulouse, France}

\author{Mathias Fink}
\affiliation{Institut Langevin, ESPCI Paris, Université PSL, CNRS, 75005 Paris, France}

\author{Valentin Mazières}
\affiliation{ISAE-SUPAERO, Université de Toulouse, France}

\date{\today}

\begin{abstract}

\end{abstract}

\maketitle

\section{Introduction}



Focusing beyond the diffraction limit is a topic of importance both fundamentally and for applications. This diffraction limit can be simply understood for the case of waves focusing in an homogeneous medium. Before focusing, a converging wave is observed, that converges towards the focusing point. This converging wave will not stop at the focusing point and will be converted into a diverging wave \cite{przadka2012time}. The focal spot that results from this interference between these convergent and divergent waves is of half the wavelength, which corresponds in fact to the well known diffraction limit \cite{ma2018towards}. This explains for example the size of the focusing spot obtained in classical time reversal (TR) experiments \cite{fink1989self,lerosey2004time}. 



Overcoming the diffraction limit, meaning focusing waves on a sub-wavelength scale, has received considerable attention for applications involving light and acoustics. Indeed, the intense focusing achieved enhances the interactions between waves and matter, and the improved spatial resolution opens up possibilities in fields such as imaging, detection and communication. Several means have been proposed to reach sub-wavelength focusing, whether or not focusing is achieved using TR.

On one hand, sub-wavelength focusing can be achieved by placing a specific medium at the focusing location. For microwaves, the spatial dimension that results from the interference between the converging and diverging waves in a random distribution of scatterers has been shown to be much smaller than half the wavelength \cite{lerosey2007focusing}. In optics, placing a nanoparticle at the focusing location can result in a flip of the sign of the diverging wave, resulting in a constructing interference with the converging wave and thus a sub-wavelength focusing \cite{noh2013broadband}.

On the other hand, a ``sink'' can be used to obtain sub-wavelength focusing. It consists in absorbing the converging wave at the focusing location so that only the converging one remains. In acoustics, two types of sink have been proposed: an active sink consisting in emitting the time-reversed source at the focusing location \cite{de2002overcoming} and a passive sink consisting in placing a near-perfect absorber at the focusing location \cite{ma2018towards}.
Impedance-matched antennas in microwaves \cite{davy2016time} and sub-wavelength metallic nanostructures in optics \cite{noh2013broadband,noh2012perfect} have also been shown to act as passive electromagnetic sinks. In all of these cases, the sink absorbs all the converging wave in order to avoid the diverging part and a sub-wavelength focusing is observed. From a fundamental point of view, a sink can be viewed as a combination of a ``reflectionless state'' \cite{sweeney2020theory} and a sub-wavelength focusing. The term ``reflectionless" applies here to the diverging wave, whose amplitude has to be equal to zero. In particular, the reflectionless state required for a passive sink has been referred to as ``coherent perfect absorption" (CPA) \cite{sweeney2020theory,chong2010coherent}, which consist of Maxwell solutions with only incoming waves at real frequencies.




In optics, a passive sink may be obtained if the incident light couples to localized surface plasmon (LSP) resonances, resulting in CPA condition \cite{noh2013broadband,noh2012perfect}. The presence of optical surface plasmons at metal interfaces is due to the angular frequency $\omega_p$ of metals, which lies in the optical regime due to their high density of free electrons \cite{pitarke2006theory}. Plasmas, which consists of ionized gazes with a lower electron density, can support surface plasmons in the microwave regime\footnote{This depends on the electron density of the considered plasma. In this paper we focus on plasmas with density resulting in a plasma angular frequency in the microwave regime.} \cite{rider2012plasmas}. Hence, we demonstrate in this paper that sub-wavelength plasmas behave as a passive microwave sink by exciting LSP resonance inside the plasma. Note that recently, it was shown that TR can concentrate an electromagnetic wave with sufficient intensity to ionize a gas and create a plasma at the focusing location \cite{mazieres2019plasma,mazieres2020spatio,mazieres2021space}. For this first study, plasma ignition will not be considered, we focus here on the linear case where a plasma is already present at the focusing location.

The aim of this paper is to develop a simulation code allowing to describe the microwave passive sink based on the excitation of LSPs resonances inside the plasma. Then this simulation code is used to study the behavior of such a sink in presence of a focusing pulse containing a broadband of frequency, similar to the pulses used in microwave TR applications. This paper is organized as follows:

In Sec. \ref{sec:plasma_sink}, the principle behind an electromagnetic sink and its application in a plasma context is presented. 

In Sec. \ref{sec:model_pres}, a model describing the plasma-based electromagnetic sink is introduced. This model involves coupling Maxwell's equations with the plasma density and current formulas. A BOR (Body of Revolution) FDTD (Finite Difference Time Domain) formalism is used to discretize these equations, enabling to describe the spatio-temporal evolution of the fields.

In Sec. \ref{sec:resultats}, the results of the simulation for a continuous wave (CW) scenario are presented, demonstrating that a plasma can indeed act as a microwave passive sink. The effectiveness of this sink is further analyzed with an incident pulse similar to the ones used in microwave TR experiments.

\section{\label{sec:plasma_sink} Principle of the Plasma Sink}

When a medium is located at the focusing spot, the total field \(\Phi_{tot}\) resulting from a converging monochromatic wave can be expressed as follows \cite{ma2018towards,noh2013broadband}:

\begin{eqnarray}
    \Phi_{tot} = G_c - s G_d
    \label{Signalpuits}
\end{eqnarray}

With, \(G_c\) and \(G_d\) the Green functions of the converging and diverging waves. The complex number \(s\) in equation (\ref{Signalpuits}) represents the phase shift and the change in amplitude of the diverging wave in contrast to the converging one.

The \(s=1\) case corresponds the focusing of waves in an homogeneous medium, where the dimension of the focal spot is of half the wavelength. The \(s=-1\) case corresponds to the transformation of a destructive interference to a constructing one at the interface vacuum-medium, resulting in a smaller focal spot and an increase of the focused field \cite{noh2013broadband}. Finally, the \(s=0\) case corresponds to the electromagnetic sink, as it translates to the absence of the diverging wave outside of the medium, as shown for example by \textit{Noh et al} in 2012 \cite{noh2012perfect}. 

\subsection{2D sink}

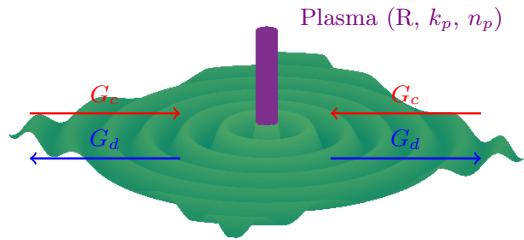
\begin{figure}[h!]
\centering
\begin{tikzpicture}
    \node at (0,0) {\includegraphics[trim=1cm 1.75cm 1cm 1.75cm, clip=true,width=7.5cm]{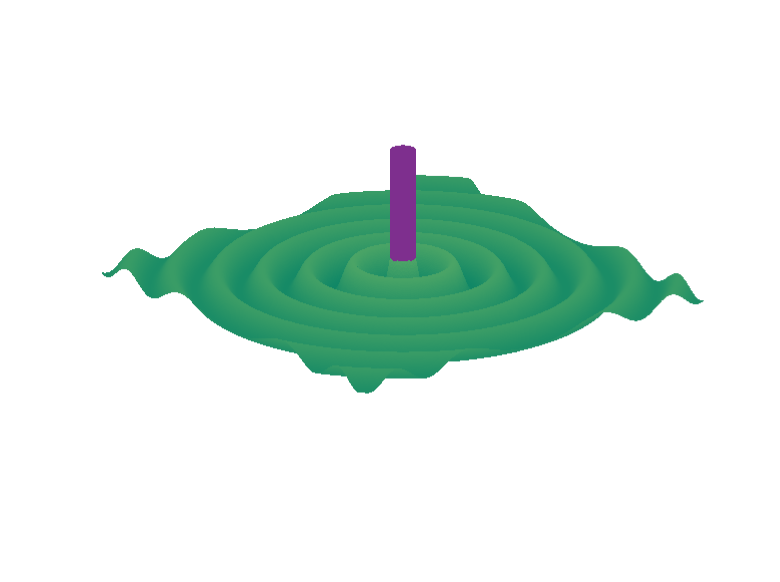}};
    \draw[->,thick,red] (-3,0.5) -- (-1,0.5) node[above,midway] {\(G_c\)};
    \draw[->,thick,red] (3,0.5) -- (1,0.5) node[above,midway] {\(G_c\)};
    \draw[->,thick,blue] (-1,-0.1) -- (-3,-0.1) node[above,midway] {\(G_d\)};
    \draw[->,thick,blue] (1,-0.1) -- (3,-0.1) node[above,midway] {\(G_d\)};
    \node[align=center,violet] at (2,1.75) { Plasma (R, \(k_p\), \(n_p\)) };
\end{tikzpicture}
\caption{\justifying{Schematic of a 2D case, in which the converging wave focuses on a cylindrical plasma.}}
\label{fig:schema2D}
\end{figure}

\begin{table}[!h]
    \caption{Different forms that the total field takes outside the medium as a function of the value of \(s\).}
    \centering 
    \begin{ruledtabular}
        \begin{tabular}{cccc}
            \(s\) & 1 & 0 & -1 \\ \hline
            \(\Phi_{tot}\) & \(J_m(k\mathbf{r})\) & \(H_m^{(2)}(k\mathbf{r})\) & \(N_m(k\mathbf{r})\) \\ 
        \end{tabular}
    \end{ruledtabular}
    \label{Psi_function_s}
\end{table}

\begin{figure*}
\setlength\unitlength{1cm}
\begin{picture}(8.6,11)
\put(-3,0){\centerline{\includegraphics[width=20cm,trim={0cm 20cm 0 1.5cm},clip]{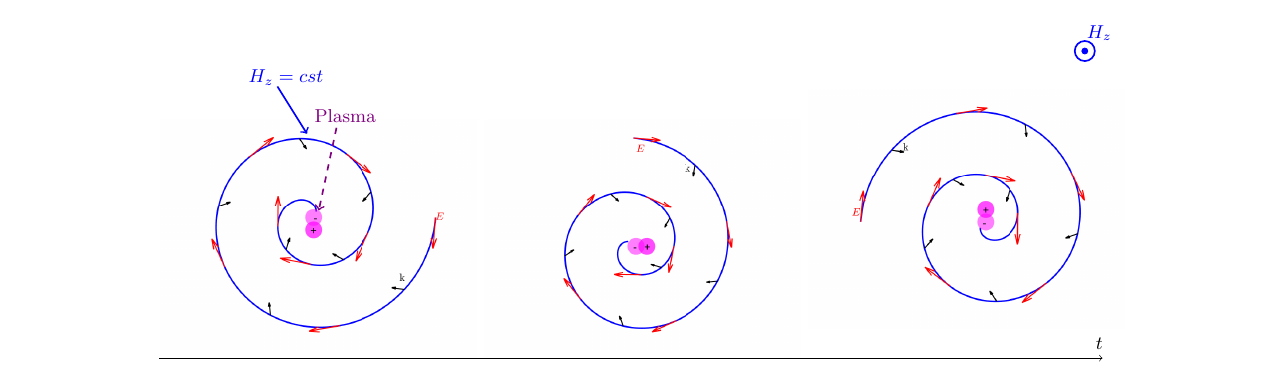}}}
\put(-3,6){\centerline{\includegraphics[width=20cm,trim={0cm 7.5cm 0 16cm},clip]{Images/Cercle_complet.pdf}}}
\put(-4.5,8){(a) $m=0$}
\put(-4.5,2.5){(b) $m=1$}
\end{picture}
\caption{Schematic representation of a converging wave for the TM case: (a) \(m=0\) and (b) \(m=1\). Each frame corresponds to a different time step (from left to right). The electric field is represented in red, the wavenumber in black and the magnetic field in blue.}\label{fig:m0_m1}
\end{figure*}

The case studied in this paper is represented in figure \ref{fig:schema2D}. It consists of a 2D converging wave \(G_c\) that converges on a plasma-cylinder of radius \(R\) (with \(R\) being subwavelength) and of an index of refraction \(n_p = \sqrt{\varepsilon_p}\) (where \(\varepsilon_p\) is the relative permittivity of the plasma). Inside the cylinder, the wave has a wavenumber \(k_p\). Outside of the cylinder, the refractive index in free-space is \(n = 1\), and the converging and diverging waves evolve with a wavenumber \(k\) in the \((r,\phi)\) plane. In 2D, the Green functions can be expressed as \(G_c(\mathbf{r},\phi) \propto H_m^{(2)}(k\mathbf{r})e^{jm\phi}\) and \(G_d(\mathbf{r},\phi) \propto H_m^{(1)}(k\mathbf{r})e^{jm\phi}\), with \(H_m^{(2)}\) and \(H_m^{(1)}\) being the Hankel functions of order \(m\) of the first and second kind, and \(k\) the wavenumber in the medium surrounding the plasma. Table \ref{Psi_function_s} shows the form that the outside field \(\Phi_{tot}\) will take in function of \(s\), with \(J_m(k\mathbf{r})\) being the Bessel function of order \(m\) and \(N_m(k\mathbf{r})\) being the Neumann function of order \(m\). From the continuity relation between the electric field \(\mathbf{E}\) and the magnetic one \(\mathbf{H}\) at the vacuum-cylinder interface (\(r = R\)) and knowing that inside the plasma the magnetic field is \(H_z(\mathbf{r},\phi) = aJ_m(k_pr)e^{jm\phi}\) (with \textit{a} a normalization constant), the \(s\) parameter can then be expressed for the transverse magnetic\footnote{The sink condition $s=0$ requires the excitation of LSPs \cite{noh2013broadband,noh2012perfect}. Since their excitation is not possible in 2D in the transverse electric case, only the TM case is considered in this work.} (TM) case as:


\begin{align}
    s = \frac{-J_m'(n_p k R) H_m^{(2)}(k R) + n_p J_m(n_p k R) H_m'^{(2)}(k R)}{J_m'(n_p k R) H_m^{(1)}(k R) - n_p J_m(n_p k R) H_m'^{(1)}(k R)}
    \label{Eq_s}
\end{align}


For a TM cylindrical wave of order $m=0$, the sink condition $s=0$ can not be reached. Indeed, for small arguments such as \(n_pkR \approx 0\), we have \(J_m(u) \underset{u\rightarrow0}{=} (u/2)^m/\Gamma(m+1)\), where \(\Gamma(m+1)\) is the Gamma function and \(u = n_pkR\) \cite{abramowitz1948handbook}. Taking this approximation into account, one can show that $s$ is always equal to 1 for $m=0$. Hence, no LSP can be excited in this case. This comes from the azimuthal symmetry of the $m=0$ wave, that results in a total electric field equal to zero at the focusing location, as illustrated in figure \ref{fig:m0_m1}(a).

For a TM cylindrical wave of higher order ($m\geq 1$), its azimuthal asymmetry  enables the presence of a nonzero electric field (in the \((r,\phi)\) plane) on the plasma cylinder. This electric field can lead to a collective movement of free electrons within the plasma and hence the excitation of LSP resonances. Figure \ref{fig:m0_m1}(b) illustrates the $m=1$ case, which is the case studied in the present paper.



\subsection{2D sink studied in this work}

To properly describe the wave in the plasma and the interaction at the interface, Maxwell's equations will be coupled with plasma equations to account for the electron current in the plasma, as in Ref. \cite{mazieres2021space}:

\begin{equation}
    \nabla \times \mathbf{E} = -\mu_0\frac{\partial \mathbf{H}}{\partial t}
    \label{M_F}
\end{equation}

\begin{equation}
    \nabla \times \mathbf{H} = -\varepsilon_0\frac{\partial \mathbf{E}}{\partial t} + \mathbf{J_p}
    \label{M_A}
\end{equation}

\begin{equation}
    \mathbf{J_p} = -en_e\mathbf{v_e}
    \label{Jp}
\end{equation}

Here, \(\varepsilon_0\) and \(\mu_0\) are the permittivity and permeability of free space, respectively. \(\mathbf{J_p}\) corresponds to the plasma current\footnote{In this paper the impact of ion in \(\mathbf{J_p}\) will not be taken into account because at microwave frequency the ion density current is far less than the electron one \cite{kourtzanidis2015adi}.}, \(n_e\) the electron density, \(e\) elementary charge of an single electron, and \(\mathbf{v_e}\) denotes the electron velocity. The simplified fluid model derived from the second moment of the Boltzmann equation is used to describe the mean electron velocity inside the plasma (with \(\nu_m\) the collision frequency and \(m_e\) the electron mass):

\begin{equation}
    \frac{\partial \mathbf{v_e}}{\partial t} = -\frac{e \mathbf{E}}{m_e} - \nu_m \mathbf{v_e}
    \label{ve}
\end{equation}

Taking the Fourier transform of equation (\ref{ve}) and injecting the expression of $\mathbf{v_e}$ into equations (\ref{Jp}), (\ref{M_F}) and (\ref{M_A}) leads to the expression of the plasma permittivity (Drude model):

\begin{eqnarray}
    \varepsilon_p = 1 - \frac{\omega_p^2}{\omega^2 + \nu_m^2} + j\frac{\nu_m}{\omega}\frac{\omega_p^2}{\omega^2 + \nu_m^2}
    \label{Drude}
\end{eqnarray}

With \(\omega_p = \sqrt{\frac{n_e e^2}{m_e \varepsilon_0}}\) representing the plasma angular frequency, \(\omega\) the wave pulsation, \(m_e\) the electron mass. The collision rate $\nu_m$ depends on the pressure $p$ and on the gaz. The plasma considered in this work is an argon plasma at low pressure, as studied in the numerical results \cite{mazieres2021space} and experimental findings \cite{mazieres2019plasma} on TR plasmas. For this gaz, the collision rate can be expressed as \(\nu_m \approx 5.4 \times 10^{9} p \text{(Torr)}\) \cite{macdonald1966microwave}.


For this study, the radius of the cylinder will be fixed at \(R=\lambda/20\), so that \(k R = 0.31\) at \(f = 2.4 \, \text{GHz}\). Theoretically, to get \(s=0\) for \(m=1\), a permittivity of \(\varepsilon_p = -1.14 + j0.17\) is needed (and for \(s = -1\), \(\varepsilon_p = -1.14\)). In the following of the paper, using equation (\ref{Drude}), electron density \(n_e\) and collision rate \(\nu_m\) are chosen to achieve the correct permittivity for both cases of \(s\), as given in Table \ref{ne_vm_bon}.

\begin{table}[!h]
    \caption{Plasma properties to achieve \(s=0\) (sink) and \(s=-1\) .}
        \begin{ruledtabular}
            \begin{tabular}{ccc}
              & \(n_e(m^{-3})\) & \(\nu_m(s^{-1})\) \\
              \hline
             \(s=-1\) & \(1.55\times 10^{17}\) & 0 \\
             \(s=0\) & \(1.55\times 10^{17}\) & \(1\times10^{9}\) \\
            \end{tabular}
        \end{ruledtabular}
    \label{ne_vm_bon}
\end{table}

\section{Numerical model}\label{sec:model_pres}


A "1.5D" BOR (Body of Revolution) FDTD (Finite Difference Time Domain) numerical model has been developed. The wave is in TM mode, the magnetic field is along the \(z-\)axis, and the electric field evolves in the \((r,\phi)\) plane. The central frequency of the converging wave \(G_c\) is of \(f_c=2.4\) GHz. The soft source of the magnetic field is placed several wavelengths away from the plasma and the total duration of the simulation is about \(40 \, \text{ns}\).




This BOR-FDTD method allows us, thanks to Maxwell's equations in cylindrical coordinates and in cases of symmetry of revolution, to eliminate the \(\phi\) dependence for the fields in the Maxwell's equation \cite{taflove2000computational}. This allows to transpose a 3D system into a ``2.5D" system that is much less resource-intensive for simulation \cite{belkhir2008extension,hadi2016radial}. In this study, the cylinder is considered semi-infinite so that the \textbf{E} and \textbf{H} fields do not depend on \(z\). This transition to ``1.5D" allows for a very thin grid despite a larger domain size, thus improving precision in the interaction at the plasma-vacuum interface. BOR-FDTD have been exploited to describe the interaction between  electromagnetic waves and plasma in toroidal tokamak geometries \cite{tierens2011time} or more recently to wave propagation in magnetized plasma \cite{wu2024vortex}. To that end, the equations describing the behavior of the plasma (equation (\ref{Jp}) and equation (\ref{ve})) must be integrated to the BOR formalism.

The discretized equations can be found in appendix \ref{ap:disc}. This numerical model is exploited in the following section to study the capability of a plasma to serve as a sink.

\section{Plasma-Based sinks}\label{sec:resultats}

This section covers the results obtained using the numerical model introduced in the previous section. To that end, electron density \(n_e\) and collision rate \(\nu_m\) are chosen to achieve the correct plasma permittivity for each case, as given in Table \ref{ne_vm_bon}. These conditions are obtained from equations (\ref{Signalpuits}) and (\ref{Drude}), meaning for a CW case. It is important to note that the numerical model presented in section \ref{sec:model_pres} enables a spatio-temporal description of the fields. Hence, the transient state of the studied CW cases is described by this model and their steady states should correspond to the conditions given by the value of the $s$ parameter. Moreover, this model enable the description of the broadband case, to analyse how the sink reacts to an incident pulse of duration \(\tau = 15\) ns, typical of pulses used in TR experiments \cite{lerosey2004time,lerosey2007focusing,mazieres2019plasma}.



\subsection{Preliminary study}

\begin{figure}[h!]
    \subfigure[]{\includegraphics[width=0.35\textwidth]{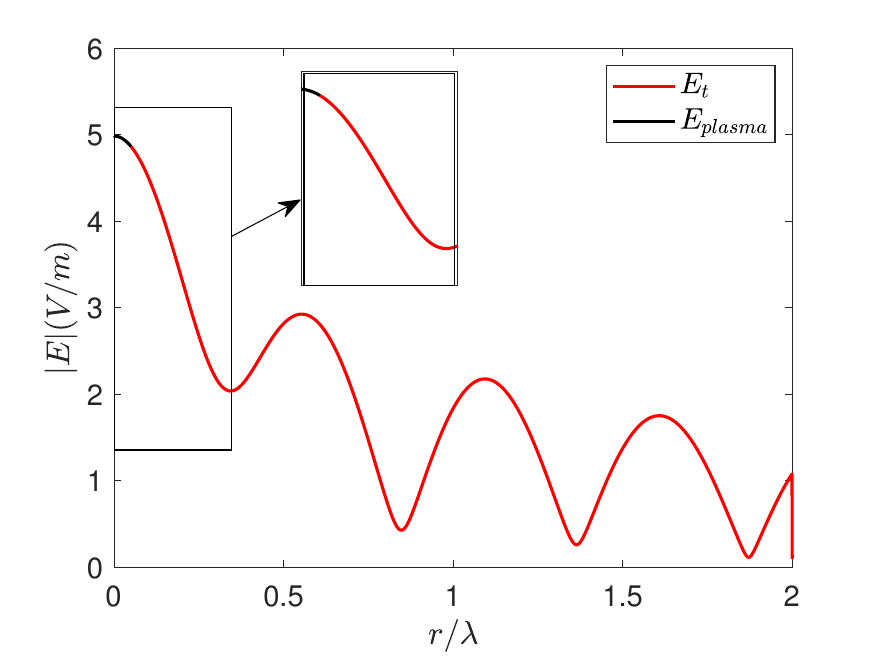}} \label{s1_r}
    \subfigure[]{\includegraphics[width=0.35\textwidth]{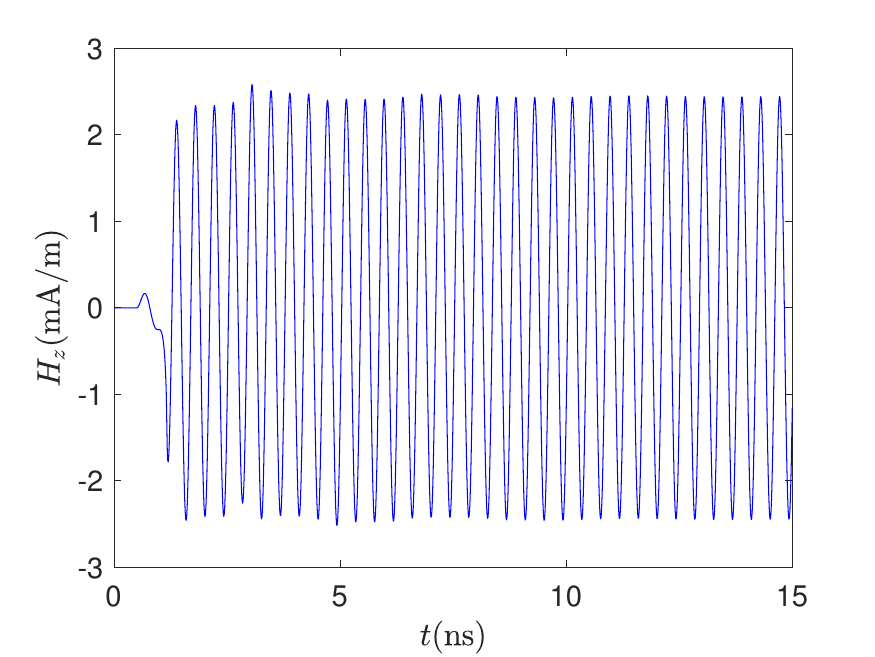}} \label{s1_time}
    \subfigure[]{\includegraphics[width=0.35\textwidth]{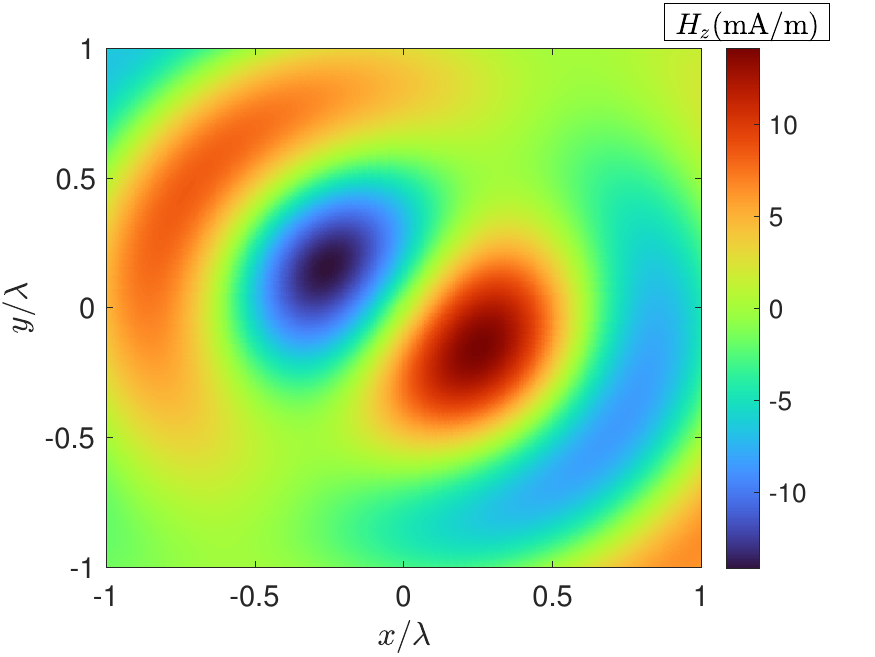}} \label{rev_s1}
    \caption{\justifying{ Case without plasma: (a) Radial distribution of \( |E| \) at \( t = 17.6\, \text{ns} \). (b)  Evolution of the diffracted magnetic field without plasma in function of time at \(r = 2\lambda\ \). (c) Revolution symmetry with \(m = 1\) of the radial distribution of \(H_z\) at \( t = 13.6\, \text{ns} \) (excerpted from movie \textcolor{red}{MEDIA 1}).}} 
    \label{fig:s1}
\end{figure}

\begin{figure}[h!]
    \subfigure[]{\includegraphics[width=0.35\textwidth]{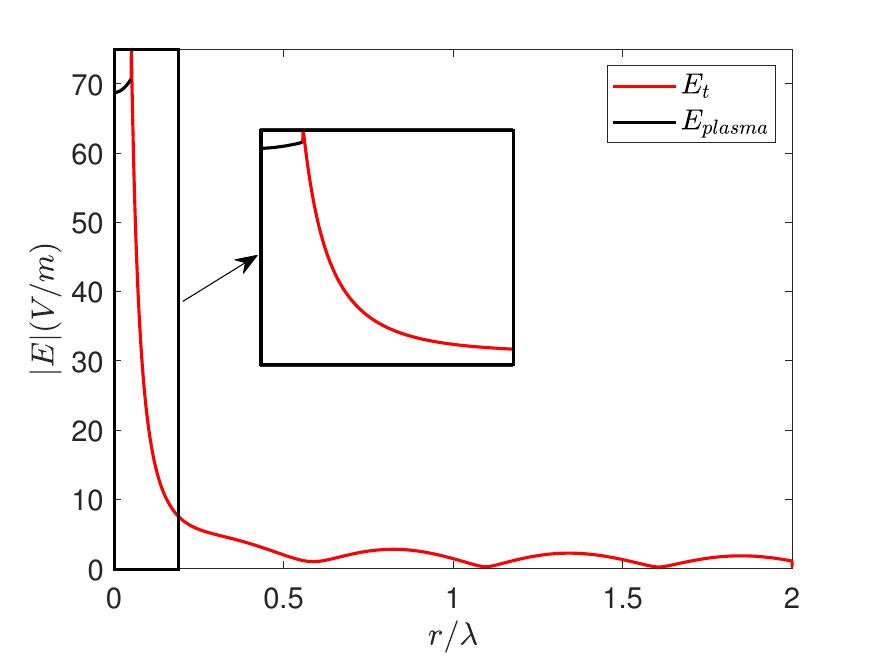}} \label{s1_r}
    \subfigure[]{\includegraphics[width=0.35\textwidth]{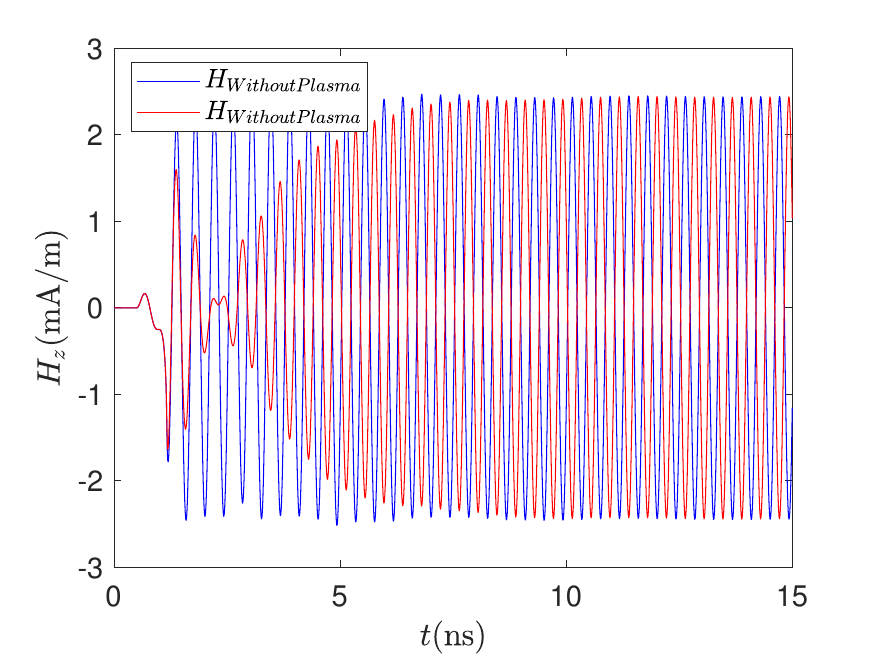}} \label{s1_time}
    \subfigure[]{\includegraphics[width=0.35\textwidth]{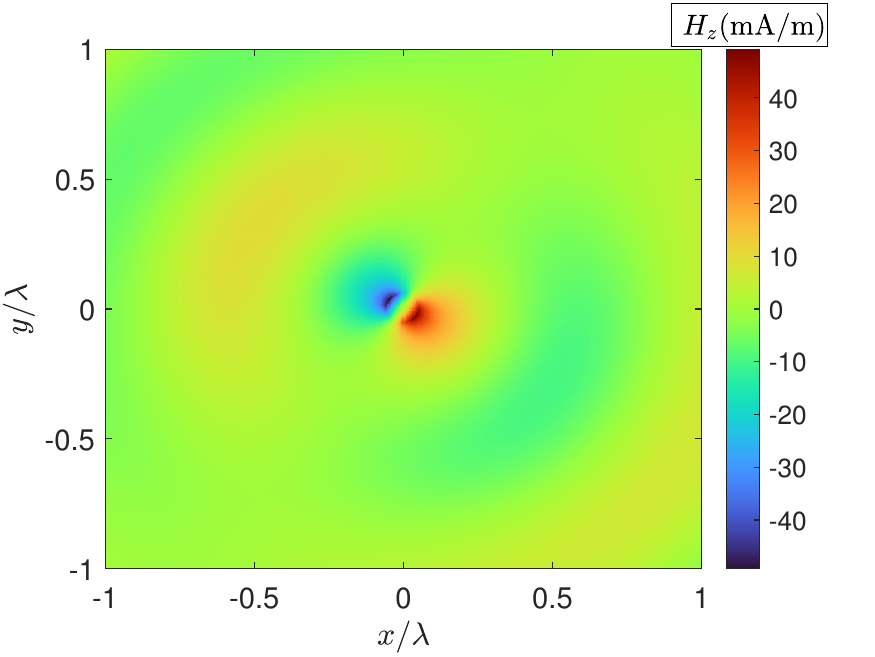}} \label{rev_s1}
    \caption{\justifying{$s=-1$: (a) Radial distribution of \( |E| \) in the steady state. (b) Evolution of the diffracted magnetic field without (blue) and with (red) plasma in function of time at \(r = 2\lambda\ \). (c) Revolution symmetry with \(m = 1\) of the radial distribution of \(H_z\) in the steady state (excerpted from movie \textcolor{red}{MEDIA 2}).}}
    \label{fig:s-1}
\end{figure}

First, let's look at what happens when there is no plasma. In such a configuration, the radial distribution, the evolution and the spatial distribution of \textbf{E} in CW is given in Figure \ref{fig:s1}. The spatial shape of the obtained field observed in Figure \ref{fig:s1}(a) comes from the considered mode (\(m=1\)). The interference created between the converging wave and the diverging wave behaves like a standing wave (as can be seen in the media). The spatial focusing dimension is found to be, as expected, \(D = \lambda/2\) and the 2D revolution shown in Figure \ref{fig:s1}(c) (excerpted from movie \textcolor{red}{MEDIA 1}) exhibit of a \(\sin(m\phi)\), typical of a standing wave resulting form the interference of the converging and diverging waves (equation (\ref{Signalpuits})). The 2D revolution is obtained by first taking the radial distribution of the field, doing his 2D revolution of it and then passing into the frequency space, multiply by \(e^{jm\phi}\), then returning to the temporal space giving such graph. The radial distribution shows a shape very similar to a Bessel function, as predicted in Table \ref{Psi_function_s}. In the following of the paper, a comparison between the diverging field without  and with plasma will allow to check the validity of the results.

If the plasma is at the theoretical conditions expressed in Table \ref{ne_vm_bon} for the \(s = -1\) case, it can be shown in Figure \ref{fig:s-1} that there is a better intensification of the focusing in the steady state, as expected. The electric field intensity increases from approximately \(|E| = 2.5 \, \text{V} \cdot \text{m}^{-1}\) in the \(s = 1\) case to \(|E| = 70 \, \text{V} \cdot \text{m}^{-1}\) for the \(s = -1\) case, as can be seen in Figure \ref{fig:s-1}(a) as well as in Figure \ref{fig:s-1}(c) (excerpted from movie \textcolor{red}{MEDIA 2}). The focalization length is \(D = \lambda/10\), which is well under the diffraction limit of \(\lambda/2\) and exhibit a \(\cos(m\phi)\) behavior. During the first 8 ns, the transient phase of the excited LSP resonance is observed, as shown in Figure \ref{fig:s-1}(b). At approximately \(4 \, \text{ns}\), the plasma does not act like a \(s = -1\) standing wave but more like a \(s = 0\) sink, absorbing part of the converging wave at the start. This behavior is typical of overcoupled resonators, for which their external decay rate is superior to the internal one (which is the case here due to the lossless nature of the plasma) \cite{delage2022experimental}. After that, the steady state of the excited LSP resonance is reached, and the wave behaves like a standing wave with the shape of a Neumann function. This steady state results aligns with what was shown previously by \textit{Noh et al.} \cite{noh2013broadband}. Under these conditions, the plasma will shift the diverging wave by \(\pi\), resulting in an out-of-phase magnetic field between the case with no plasma (in blue on  Figure \ref{fig:s-1}(b)) and the case with plasma (in red on  Figure \ref{fig:s-1}(b)). It means that when the wave exits the plasma, the diverging wave will be in phase with the converging wave, creating constructive interference just as predicted in Table \ref{Psi_function_s} for the case of \(s = -1\). This is made possible by the excitation of LSP resonance, as highlighted by the intense magnetic field obtained at the plasma-vacuum interface. The fact that the wave exhibits a Neumann-type shape while the field magnitude does not diverge stems from the observation that the field equals a Neumann function only outside of the plasma. Since the plasma has a finite radius, the electric field magnitude cannot diverge and has its maximum intensity at \(r = R\) \cite{noh2013broadband}.

\subsection{\label{sec:CW} CW plasma-based sink}

\begin{figure}[h!]
    \subfigure[]{\includegraphics[width=0.35\textwidth]{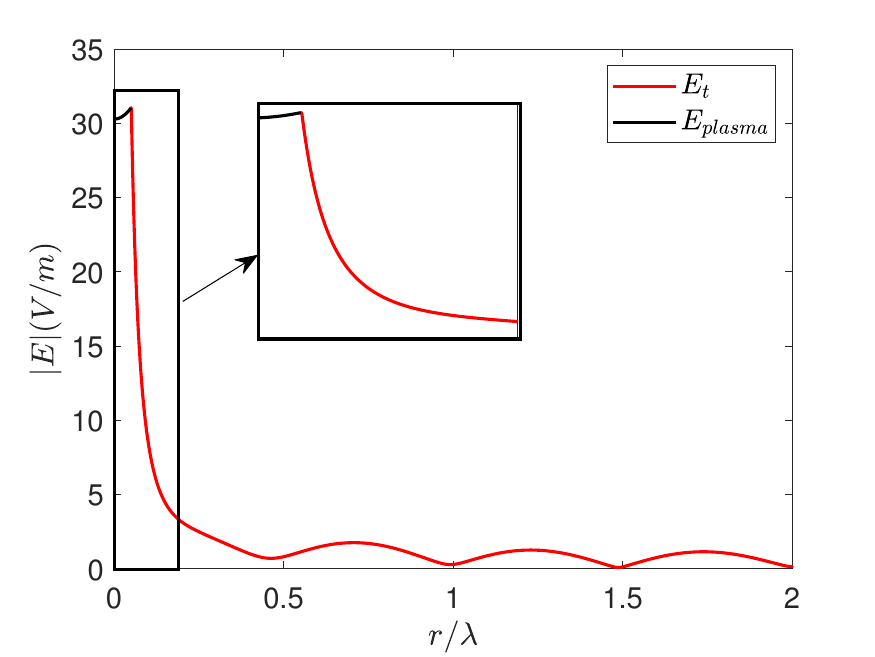}} \label{s1_r}
    \subfigure[]{\includegraphics[width=0.35\textwidth]{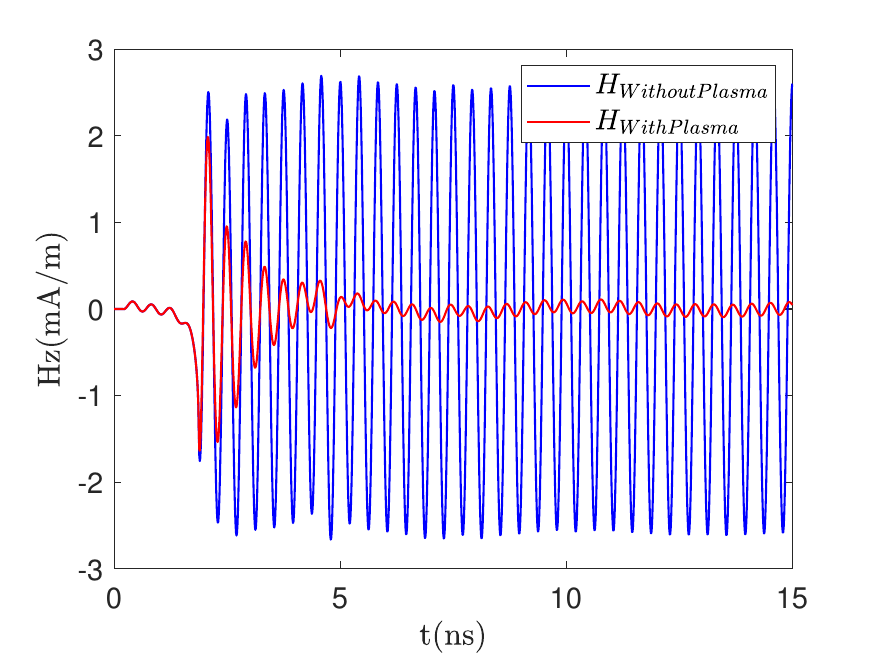}} \label{s1_time}
    \subfigure[]{\includegraphics[width=0.35\textwidth]{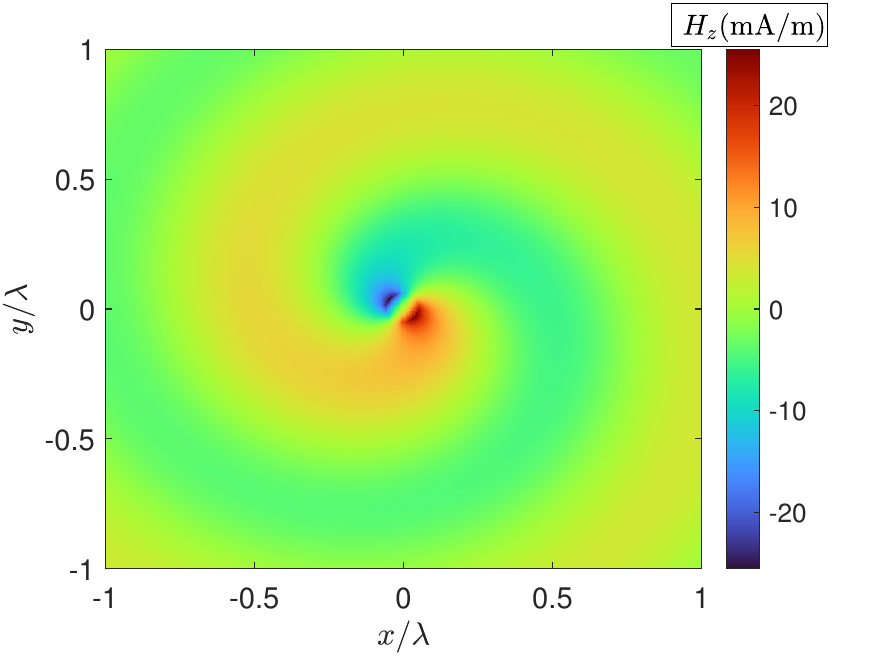}} \label{rev_s1}
    \caption{\justifying{$s=0$: (a) Radial distribution of \( |E| \) in the steady state. (b)Evolution of the diffracted magnetic field without (blue) and with (red) plasma in function of time at \(r = 2\lambda\ \). (c) Revolution symmetry with \(m = 1\) of the radial distribution of \(H_z\) in the steady state (excerpted from movie \textcolor{red}{MEDIA 3}).}}
    \label{fig:s0}
\end{figure}

The plasma is now in the theoretical conditions as stated in Table \ref{ne_vm_bon} for the \(s = 0\) case, namely \(n_e = 1.55\times 10^{17} \text{m}^{-3}\) and  \(\nu_m = 10^9 \text{s}^{-1}\). Figure \ref{fig:s0}(a) shows, in the steady state, a focalization length of approximately \(D = \lambda/10\) and an intensification of the focalization point compared to the \(s = 1\) case. However, the intensification is not as great as in the \(s = -1\) case, due to dissipation in the plasma ($\nu_m>0$). The transient regime is also observed in Figure \ref{fig:s0}(b), reflecting the excitation of the LSP resonance. Once the steady state is reached, the diverging field outside the plasma is nearly equal to zero, meaning the great majority of the converging wave is absorbed by the plasma. The 2D revolution of Figure \ref{fig:s0}(c) (excerpted from movie \textcolor{red}{MEDIA 3}) shows a \(e^{jm\phi}\), implying having only a converging wave, similarly to what \textit{Noh et al.} \cite{noh2012perfect} found. The spiral shape of the magnetic field is clearly visible, corresponding to the $m=1$ case illustrated in figure \ref{fig:m0_m1}(b). Moreover, the intense magnetic fields obtained at the plasma-vacuum interface is due to the excitation of a LSP resonance, illustrated in  figure \ref{fig:m0_m1}(b). Thus, the wave tends to behave more like a Hankel function. The progressive nature of this wave, reflecting the absence of the diverging wave, can be seen in movie \textcolor{red}{MEDIA 3} (the different steps shown in figure \ref{fig:m0_m1}(b) are also found in this movie). It's magnitude seems to diverge as \(r\) approaches zero (the field magnitude does not diverge for the same reason as for \(s = -1\)), giving rise to the intensification and reduction in focalization length.

\subsection{\label{sec:broadband} Broadband plasma-based sink}

\begin{figure}[h!]
    \subfigure[]{\includegraphics[width=0.35\textwidth]{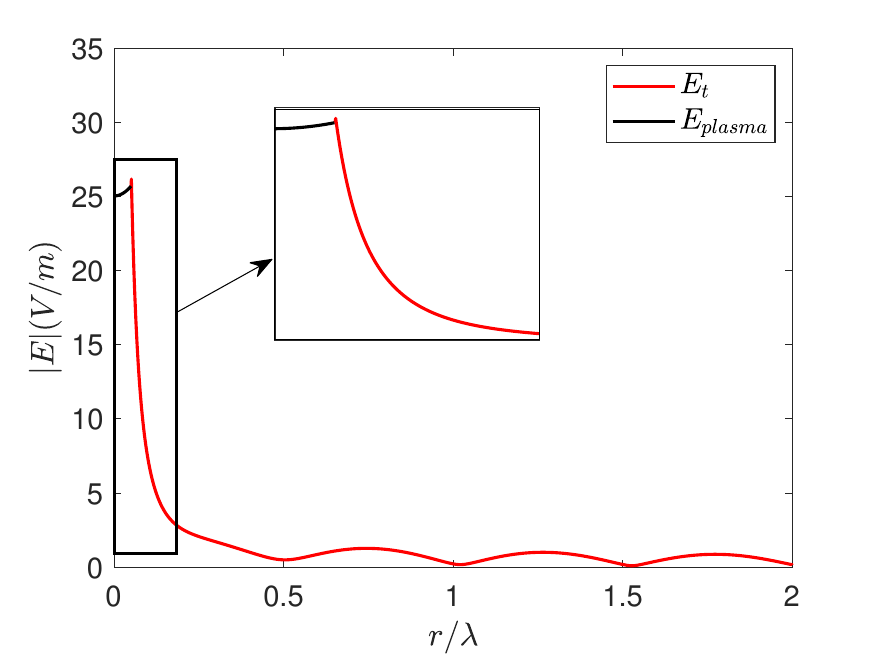}} 
    \subfigure[]{\includegraphics[width=0.35\textwidth]{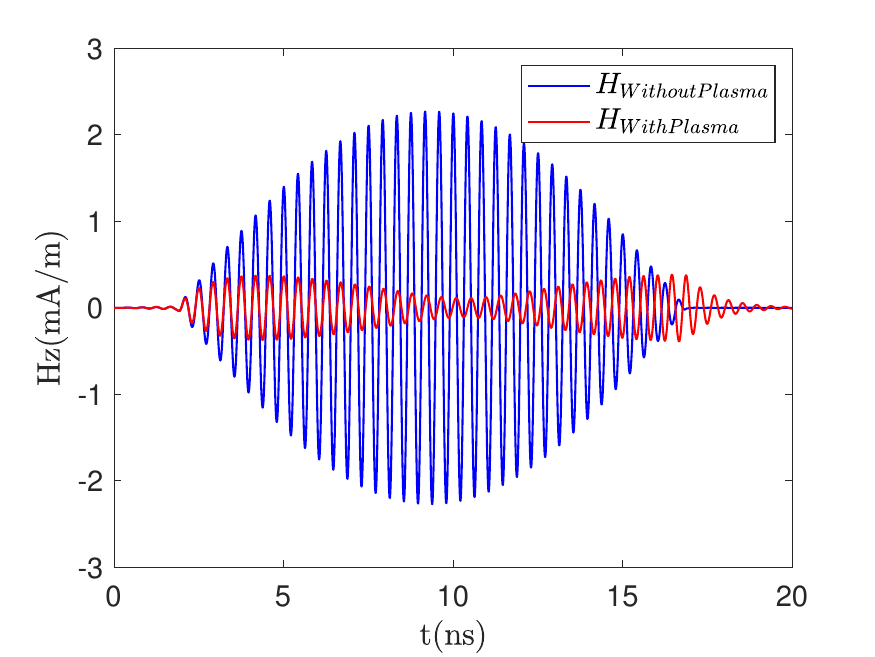}} 
    \subfigure[]{\includegraphics[width=0.35\textwidth]{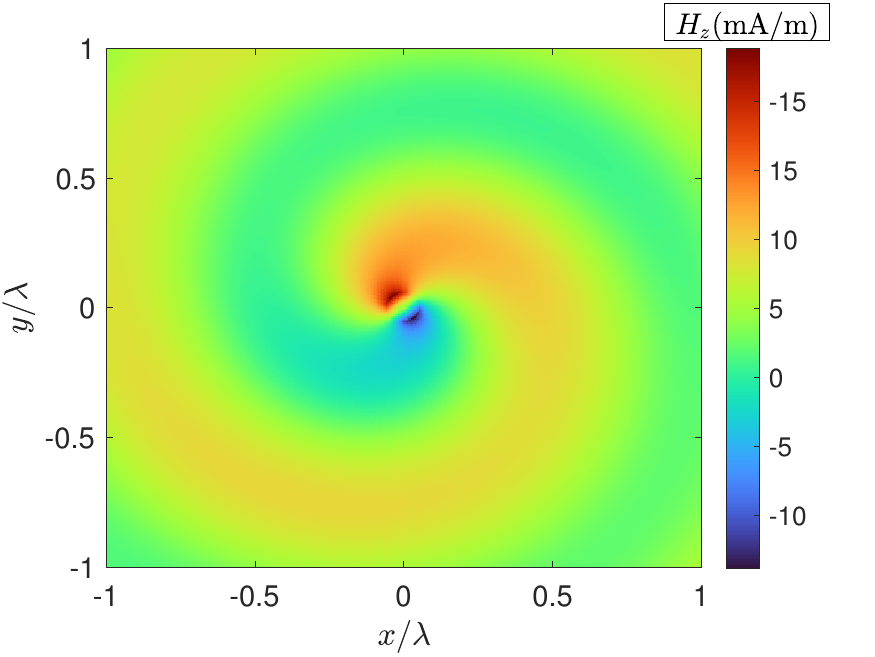}} 
    \caption{\justifying{$s=0$: (a) Radial distribution of the \(\mathbf{E}\) fields with a pulse (\(\tau = 15\, \text{ns}\)) at its maximum. (b) Evolution of the diffracted magnetic field without (blue) and with (red) plasma in function of time at \(r = 2\lambda\). (c) Revolution symmetry with \(m = 1\) of the radial distribution at its maximum.}} 
    \label{fig:Ei_Es_pulse_15ns}
\end{figure}

\begin{figure}[h!]
    \includegraphics[width=8cm]{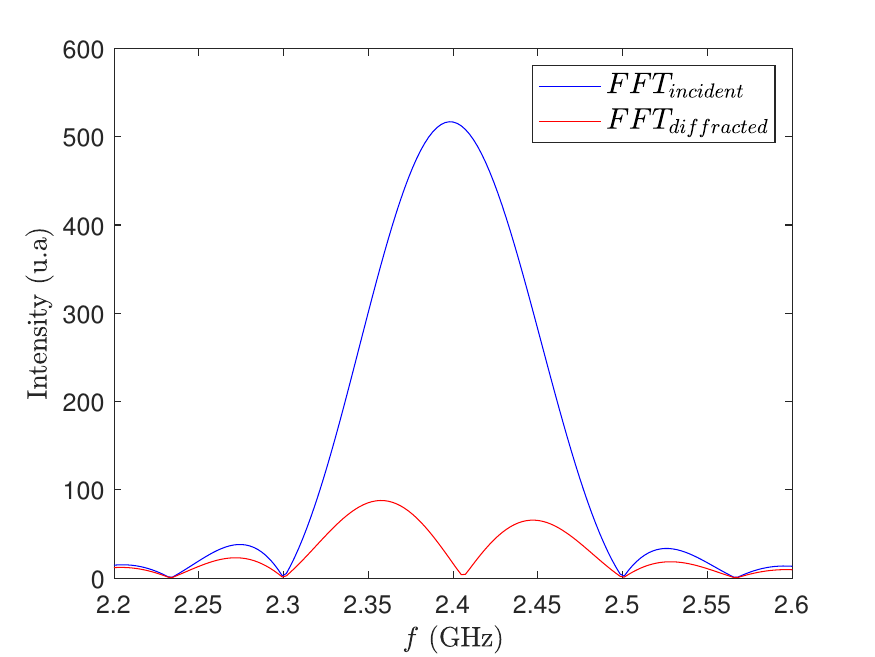}
    \caption{\justifying{Norm of the Fast Fourier Transform of the convergent (blue) and diffracted (red) pulses of figure \ref{fig:Ei_Es_pulse_15ns}(b). }}
    \label{fig:FFT_pulse_15ns}
\end{figure}

In this section the \(s=0\) case for the converging pulse of figure \ref{fig:Ei_Es_pulse_15ns}(b) is considered. This pulse of duration $\tau=15$ ns contains a broadband of frequencies as can be seen in figure \ref{fig:FFT_pulse_15ns}. This duration corresponds to the typical duration with which plasmas have been ignited by TR \cite{mazieres2019plasma,mazieres2020spatio,mazieres2021space}. Plasma being a dispersive medium, each frequency component of the converging pulse will interact differently with it. As a consequence, the value of $s$ depends strongly on the frequency. So like \textit{Noh et al} \cite{noh2013broadband} or \textit{Kim et al} \cite{kim2016general} explain, it's important to choose wisely the properties of the medium (here \(n_e\) and \(\nu_m\) for a plasma), so that every frequencies of the converging wave give the minimal value for \(s\). In our case the values \(n_e = 1.55\times10^{17}m^{-3}\) and \(\nu_m = 1.23 \times 10^9 \, \text{s}^{-1}\) are chosen so that \(s\) is small for the majority of the frequency in the bandwith, corresponding to the $s=0$ case for a frequency of 2.4 GHz.

Figure \ref{fig:Ei_Es_pulse_15ns} shows, first that the radial distribution and the 2D revolution shows a similar behavior to the one of Figure \ref{fig:s0}. In particular, an intensification of the field and a sub-wavelength focusing are present, meaning the case \(s=0\) seems to be achieved. This is possible because the duration of the pulse is higher than the one of the transient state identified in section \ref{sec:CW} (duration of approximately 3 ns, as can be seen in figure \ref{fig:s0}(b)). On the plots of the diverging pulse with and without the plasma, it can be noted that the intensity of the diverging wave with plasma is clearly diminished in contrast of the one without plasma. However, the amplitude of the field with plasma can reach about 15 \% of the maximum value of the converging pulse. To understand why this happens, the Fourier transforms of these pulses are plotted on Figure \ref{fig:FFT_pulse_15ns}. It is clear that the frequency component of the converging pulse at 2.4 GHz is well-absorbed by the plasma sink. However, certain of the other frequency components are still present on the diverging pulse, that results in the relatively low diverging pulse of Figure \ref{fig:Ei_Es_pulse_15ns}(b). This comes from the sensitivity of the CPA condition, that requires the excitation of LSP resonance.

\subsection{Discussion}

These results confirm the theoretical prediction of section \ref{sec:plasma_sink} that a plasma can behave as an electromagnetic sink at microwave frequency. In practice, cylindrical wave of order $m\geq 1$ can be generated with circular array of monopole antennas \cite{islam2023generation}. Furthermore, a plasma should be ignited inside of a cylindrical tube with the desired density ($\approx 10^{17}$ m$^{-3}$) at the desired pressure (0.2 torr). This seems to be achievable experimentally. Indeed, cylindrical tubes have been widely used to generate glow discharges in pressure ranges close or containing the desired pressure \cite{hirsh2012gaseous,fu2017investigation,pugliese2015complex,pugliese2015complex,chiad2009construction}, with maximal densities up to $10^{19}$ m$^{-3}$ \cite{pugliese2015complex,chiad2009construction}. With the small diameter ($\approx 1$ cm) required for the plasma tube, stable glow discharges can be obtained at 0.2 torr for direct current of about 50 mA \cite{hirsh2012gaseous}. Focusing cylindrical electromagnetic waves of central frequency of 2.4 GHz on a plasma tube should lead to CPA condition through the excitation of LSP resonances, and thus to an enhanced electric field and sub-wavelength focusing. 








\section{\label{sec:conclusion} Conclusion}

In the present paper, we have presented the first study on the potential use of plasmas as electromagnetic sinks. The idea is to exploit LSP resonances to reach CPA condition, as done in optics for metallic nanostructures. 

First, the theoretical conditions, meaning the plasma density and pressure required, are identified. Then, a BOR-FDTD numerical model, allowing to describe plasma-wave interaction, is introduced. This model enables to describe the spatio-temporal evolution of the fields and thus the description of the transient and steady regimes, as well as the spatial focusing. Finally, the theoretical predictions are verified with this numerical model. It is shown that the plasma can indeed behave as an electromagnetic sink, for a CW or a broadband converging wave. 

Experimentally, the plasma conditions seem reachable, with a plasma ignited inside a cylindrical tube. Focusing waves on this plasma tube could then allow to enhance plasma-wave interaction.
First, the properties of the sink to prevent the creation of divergent waves could be useful for plasma cloaking. The use of metamaterials as cloaking systems has already been studied \cite{fleury2015invisibility}, and more recently, it has been shown that plasma could be used as a cloaking system \cite{sakai2016plasma}, but it requires a specific \(n_e\) distribution. The fact that the sink requires a homogeneous \(n_e\) to work could be a promising alternative for this application. 

The intensification of the field and the sub-wavelength nature of the focusing could also enable to enhance nonlinear plasma ignition. To account for these phenomenon, nonlinear behavior of the plasma should be incorporated into the model. Further work is planned on this concept in the future. The results of the present paper represent the first essential step that needed to be achieved: showing that a plasma can behave as a microwave sink.


\newpage
~
\newpage
\bibliography{apssamp}

\begin{thebibliography}{34}%
\makeatletter
\providecommand \@ifxundefined [1]{%
 \@ifx{#1\undefined}
}%
\providecommand \@ifnum [1]{%
 \ifnum #1\expandafter \@firstoftwo
 \else \expandafter \@secondoftwo
 \fi
}%
\providecommand \@ifx [1]{%
 \ifx #1\expandafter \@firstoftwo
 \else \expandafter \@secondoftwo
 \fi
}%
\providecommand \natexlab [1]{#1}%
\providecommand \enquote  [1]{``#1''}%
\providecommand \bibnamefont  [1]{#1}%
\providecommand \bibfnamefont [1]{#1}%
\providecommand \citenamefont [1]{#1}%
\providecommand \href@noop [0]{\@secondoftwo}%
\providecommand \href [0]{\begingroup \@sanitize@url \@href}%
\providecommand \@href[1]{\@@startlink{#1}\@@href}%
\providecommand \@@href[1]{\endgroup#1\@@endlink}%
\providecommand \@sanitize@url [0]{\catcode `\\12\catcode `\$12\catcode `\&12\catcode `\#12\catcode `\^12\catcode `\_12\catcode `\%12\relax}%
\providecommand \@@startlink[1]{}%
\providecommand \@@endlink[0]{}%
\providecommand \url  [0]{\begingroup\@sanitize@url \@url }%
\providecommand \@url [1]{\endgroup\@href {#1}{\urlprefix }}%
\providecommand \urlprefix  [0]{URL }%
\providecommand \Eprint [0]{\href }%
\providecommand \doibase [0]{https://doi.org/}%
\providecommand \selectlanguage [0]{\@gobble}%
\providecommand \bibinfo  [0]{\@secondoftwo}%
\providecommand \bibfield  [0]{\@secondoftwo}%
\providecommand \translation [1]{[#1]}%
\providecommand \BibitemOpen [0]{}%
\providecommand \bibitemStop [0]{}%
\providecommand \bibitemNoStop [0]{.\EOS\space}%
\providecommand \EOS [0]{\spacefactor3000\relax}%
\providecommand \BibitemShut  [1]{\csname bibitem#1\endcsname}%
\let\auto@bib@innerbib\@empty
\bibitem [{\citenamefont {Przadka}\ \emph {et~al.}(2012)\citenamefont {Przadka}, \citenamefont {Feat}, \citenamefont {Petitjeans}, \citenamefont {Pagneux}, \citenamefont {Maurel},\ and\ \citenamefont {Fink}}]{przadka2012time}%
  \BibitemOpen
  \bibfield  {author} {\bibinfo {author} {\bibfnamefont {A.}~\bibnamefont {Przadka}}, \bibinfo {author} {\bibfnamefont {S.}~\bibnamefont {Feat}}, \bibinfo {author} {\bibfnamefont {P.}~\bibnamefont {Petitjeans}}, \bibinfo {author} {\bibfnamefont {V.}~\bibnamefont {Pagneux}}, \bibinfo {author} {\bibfnamefont {A.}~\bibnamefont {Maurel}},\ and\ \bibinfo {author} {\bibfnamefont {M.}~\bibnamefont {Fink}},\ }\bibfield  {title} {\bibinfo {title} {Time reversal of water waves},\ }\href@noop {} {\bibfield  {journal} {\bibinfo  {journal} {Physical review letters}\ }\textbf {\bibinfo {volume} {109}},\ \bibinfo {pages} {064501} (\bibinfo {year} {2012})}\BibitemShut {NoStop}%
\bibitem [{\citenamefont {Ma}\ \emph {et~al.}(2018)\citenamefont {Ma}, \citenamefont {Fan}, \citenamefont {Ma}, \citenamefont {De~Rosny}, \citenamefont {Sheng},\ and\ \citenamefont {Fink}}]{ma2018towards}%
  \BibitemOpen
  \bibfield  {author} {\bibinfo {author} {\bibfnamefont {G.}~\bibnamefont {Ma}}, \bibinfo {author} {\bibfnamefont {X.}~\bibnamefont {Fan}}, \bibinfo {author} {\bibfnamefont {F.}~\bibnamefont {Ma}}, \bibinfo {author} {\bibfnamefont {J.}~\bibnamefont {De~Rosny}}, \bibinfo {author} {\bibfnamefont {P.}~\bibnamefont {Sheng}},\ and\ \bibinfo {author} {\bibfnamefont {M.}~\bibnamefont {Fink}},\ }\bibfield  {title} {\bibinfo {title} {Towards anti-causal green’s function for three-dimensional sub-diffraction focusing},\ }\href@noop {} {\bibfield  {journal} {\bibinfo  {journal} {Nature Physics}\ }\textbf {\bibinfo {volume} {14}},\ \bibinfo {pages} {608} (\bibinfo {year} {2018})}\BibitemShut {NoStop}%
\bibitem [{\citenamefont {Fink}\ \emph {et~al.}(1989)\citenamefont {Fink}, \citenamefont {Prada}, \citenamefont {Wu},\ and\ \citenamefont {Cassereau}}]{fink1989self}%
  \BibitemOpen
  \bibfield  {author} {\bibinfo {author} {\bibfnamefont {M.}~\bibnamefont {Fink}}, \bibinfo {author} {\bibfnamefont {C.}~\bibnamefont {Prada}}, \bibinfo {author} {\bibfnamefont {F.}~\bibnamefont {Wu}},\ and\ \bibinfo {author} {\bibfnamefont {D.}~\bibnamefont {Cassereau}},\ }\bibfield  {title} {\bibinfo {title} {Self focusing in inhomogeneous media with time reversal acoustic mirrors},\ }in\ \href@noop {} {\emph {\bibinfo {booktitle} {Proceedings., IEEE Ultrasonics Symposium,}}}\ (\bibinfo {organization} {IEEE},\ \bibinfo {year} {1989})\ pp.\ \bibinfo {pages} {681--686}\BibitemShut {NoStop}%
\bibitem [{\citenamefont {Lerosey}\ \emph {et~al.}(2004)\citenamefont {Lerosey}, \citenamefont {de~Rosny}, \citenamefont {Tourin}, \citenamefont {Derode}, \citenamefont {Montaldo},\ and\ \citenamefont {Fink}}]{lerosey2004time}%
  \BibitemOpen
  \bibfield  {author} {\bibinfo {author} {\bibfnamefont {G.}~\bibnamefont {Lerosey}}, \bibinfo {author} {\bibfnamefont {J.}~\bibnamefont {de~Rosny}}, \bibinfo {author} {\bibfnamefont {A.}~\bibnamefont {Tourin}}, \bibinfo {author} {\bibfnamefont {A.}~\bibnamefont {Derode}}, \bibinfo {author} {\bibfnamefont {G.}~\bibnamefont {Montaldo}},\ and\ \bibinfo {author} {\bibfnamefont {M.}~\bibnamefont {Fink}},\ }\bibfield  {title} {\bibinfo {title} {Time reversal of electromagnetic waves},\ }\href@noop {} {\bibfield  {journal} {\bibinfo  {journal} {Physical review letters}\ }\textbf {\bibinfo {volume} {92}},\ \bibinfo {pages} {193904} (\bibinfo {year} {2004})}\BibitemShut {NoStop}%
\bibitem [{\citenamefont {Lerosey}\ \emph {et~al.}(2007)\citenamefont {Lerosey}, \citenamefont {De~Rosny}, \citenamefont {Tourin},\ and\ \citenamefont {Fink}}]{lerosey2007focusing}%
  \BibitemOpen
  \bibfield  {author} {\bibinfo {author} {\bibfnamefont {G.}~\bibnamefont {Lerosey}}, \bibinfo {author} {\bibfnamefont {J.}~\bibnamefont {De~Rosny}}, \bibinfo {author} {\bibfnamefont {A.}~\bibnamefont {Tourin}},\ and\ \bibinfo {author} {\bibfnamefont {M.}~\bibnamefont {Fink}},\ }\bibfield  {title} {\bibinfo {title} {Focusing beyond the diffraction limit with far-field time reversal},\ }\href@noop {} {\bibfield  {journal} {\bibinfo  {journal} {Science}\ }\textbf {\bibinfo {volume} {315}},\ \bibinfo {pages} {1120} (\bibinfo {year} {2007})}\BibitemShut {NoStop}%
\bibitem [{\citenamefont {Noh}\ \emph {et~al.}(2013)\citenamefont {Noh}, \citenamefont {Popoff},\ and\ \citenamefont {Cao}}]{noh2013broadband}%
  \BibitemOpen
  \bibfield  {author} {\bibinfo {author} {\bibfnamefont {H.}~\bibnamefont {Noh}}, \bibinfo {author} {\bibfnamefont {S.~M.}\ \bibnamefont {Popoff}},\ and\ \bibinfo {author} {\bibfnamefont {H.}~\bibnamefont {Cao}},\ }\bibfield  {title} {\bibinfo {title} {Broadband subwavelength focusing of light using a passive sink},\ }\href@noop {} {\bibfield  {journal} {\bibinfo  {journal} {Optics Express}\ }\textbf {\bibinfo {volume} {21}},\ \bibinfo {pages} {17435} (\bibinfo {year} {2013})}\BibitemShut {NoStop}%
\bibitem [{\citenamefont {de~Rosny}\ and\ \citenamefont {Fink}(2002)}]{de2002overcoming}%
  \BibitemOpen
  \bibfield  {author} {\bibinfo {author} {\bibfnamefont {J.}~\bibnamefont {de~Rosny}}\ and\ \bibinfo {author} {\bibfnamefont {M.}~\bibnamefont {Fink}},\ }\bibfield  {title} {\bibinfo {title} {Overcoming the diffraction limit in wave physics using a time-reversal mirror and a novel acoustic sink},\ }\href@noop {} {\bibfield  {journal} {\bibinfo  {journal} {Physical review letters}\ }\textbf {\bibinfo {volume} {89}},\ \bibinfo {pages} {124301} (\bibinfo {year} {2002})}\BibitemShut {NoStop}%
\bibitem [{\citenamefont {Davy}\ \emph {et~al.}(2016)\citenamefont {Davy}, \citenamefont {de~Rosny},\ and\ \citenamefont {Besnier}}]{davy2016time}%
  \BibitemOpen
  \bibfield  {author} {\bibinfo {author} {\bibfnamefont {M.}~\bibnamefont {Davy}}, \bibinfo {author} {\bibfnamefont {J.}~\bibnamefont {de~Rosny}},\ and\ \bibinfo {author} {\bibfnamefont {P.}~\bibnamefont {Besnier}},\ }\bibfield  {title} {\bibinfo {title} {Time reversal with absorbing antennas in a mode-stirred reveberation chamber},\ }in\ \href@noop {} {\emph {\bibinfo {booktitle} {2016 IEEE Metrology for Aerospace (MetroAeroSpace)}}}\ (\bibinfo {organization} {IEEE},\ \bibinfo {year} {2016})\ pp.\ \bibinfo {pages} {156--160}\BibitemShut {NoStop}%
\bibitem [{\citenamefont {Noh}\ \emph {et~al.}(2012)\citenamefont {Noh}, \citenamefont {Chong}, \citenamefont {Stone},\ and\ \citenamefont {Cao}}]{noh2012perfect}%
  \BibitemOpen
  \bibfield  {author} {\bibinfo {author} {\bibfnamefont {H.}~\bibnamefont {Noh}}, \bibinfo {author} {\bibfnamefont {Y.}~\bibnamefont {Chong}}, \bibinfo {author} {\bibfnamefont {A.~D.}\ \bibnamefont {Stone}},\ and\ \bibinfo {author} {\bibfnamefont {H.}~\bibnamefont {Cao}},\ }\bibfield  {title} {\bibinfo {title} {Perfect coupling of light to surface plasmons by coherent absorption},\ }\href@noop {} {\bibfield  {journal} {\bibinfo  {journal} {Physical review letters}\ }\textbf {\bibinfo {volume} {108}},\ \bibinfo {pages} {186805} (\bibinfo {year} {2012})}\BibitemShut {NoStop}%
\bibitem [{\citenamefont {Sweeney}\ \emph {et~al.}(2020)\citenamefont {Sweeney}, \citenamefont {Hsu},\ and\ \citenamefont {Stone}}]{sweeney2020theory}%
  \BibitemOpen
  \bibfield  {author} {\bibinfo {author} {\bibfnamefont {W.~R.}\ \bibnamefont {Sweeney}}, \bibinfo {author} {\bibfnamefont {C.~W.}\ \bibnamefont {Hsu}},\ and\ \bibinfo {author} {\bibfnamefont {A.~D.}\ \bibnamefont {Stone}},\ }\bibfield  {title} {\bibinfo {title} {Theory of reflectionless scattering modes},\ }\href@noop {} {\bibfield  {journal} {\bibinfo  {journal} {Physical Review A}\ }\textbf {\bibinfo {volume} {102}},\ \bibinfo {pages} {063511} (\bibinfo {year} {2020})}\BibitemShut {NoStop}%
\bibitem [{\citenamefont {Chong}\ \emph {et~al.}(2010)\citenamefont {Chong}, \citenamefont {Ge}, \citenamefont {Cao},\ and\ \citenamefont {Stone}}]{chong2010coherent}%
  \BibitemOpen
  \bibfield  {author} {\bibinfo {author} {\bibfnamefont {Y.}~\bibnamefont {Chong}}, \bibinfo {author} {\bibfnamefont {L.}~\bibnamefont {Ge}}, \bibinfo {author} {\bibfnamefont {H.}~\bibnamefont {Cao}},\ and\ \bibinfo {author} {\bibfnamefont {A.~D.}\ \bibnamefont {Stone}},\ }\bibfield  {title} {\bibinfo {title} {Coherent perfect absorbers: time-reversed lasers},\ }\href@noop {} {\bibfield  {journal} {\bibinfo  {journal} {Physical review letters}\ }\textbf {\bibinfo {volume} {105}},\ \bibinfo {pages} {053901} (\bibinfo {year} {2010})}\BibitemShut {NoStop}%
\bibitem [{\citenamefont {Pitarke}\ \emph {et~al.}(2006)\citenamefont {Pitarke}, \citenamefont {Silkin}, \citenamefont {Chulkov},\ and\ \citenamefont {Echenique}}]{pitarke2006theory}%
  \BibitemOpen
  \bibfield  {author} {\bibinfo {author} {\bibfnamefont {J.}~\bibnamefont {Pitarke}}, \bibinfo {author} {\bibfnamefont {V.}~\bibnamefont {Silkin}}, \bibinfo {author} {\bibfnamefont {E.}~\bibnamefont {Chulkov}},\ and\ \bibinfo {author} {\bibfnamefont {P.}~\bibnamefont {Echenique}},\ }\bibfield  {title} {\bibinfo {title} {Theory of surface plasmons and surface-plasmon polaritons},\ }\href@noop {} {\bibfield  {journal} {\bibinfo  {journal} {Reports on progress in physics}\ }\textbf {\bibinfo {volume} {70}},\ \bibinfo {pages} {1} (\bibinfo {year} {2006})}\BibitemShut {NoStop}%
\bibitem [{\citenamefont {Rider}\ \emph {et~al.}(2012)\citenamefont {Rider}, \citenamefont {Ostrikov},\ and\ \citenamefont {Furman}}]{rider2012plasmas}%
  \BibitemOpen
  \bibfield  {author} {\bibinfo {author} {\bibfnamefont {A.}~\bibnamefont {Rider}}, \bibinfo {author} {\bibfnamefont {K.}~\bibnamefont {Ostrikov}},\ and\ \bibinfo {author} {\bibfnamefont {S.}~\bibnamefont {Furman}},\ }\bibfield  {title} {\bibinfo {title} {Plasmas meet plasmonics: Everything old is new again},\ }\href@noop {} {\bibfield  {journal} {\bibinfo  {journal} {The European Physical Journal D}\ }\textbf {\bibinfo {volume} {66}},\ \bibinfo {pages} {1} (\bibinfo {year} {2012})}\BibitemShut {NoStop}%
\bibitem [{\citenamefont {Mazieres}\ \emph {et~al.}(2019)\citenamefont {Mazieres}, \citenamefont {Pascaud}, \citenamefont {Liard}, \citenamefont {Dap}, \citenamefont {Clergereaux},\ and\ \citenamefont {Pascal}}]{mazieres2019plasma}%
  \BibitemOpen
  \bibfield  {author} {\bibinfo {author} {\bibfnamefont {V.}~\bibnamefont {Mazieres}}, \bibinfo {author} {\bibfnamefont {R.}~\bibnamefont {Pascaud}}, \bibinfo {author} {\bibfnamefont {L.}~\bibnamefont {Liard}}, \bibinfo {author} {\bibfnamefont {S.}~\bibnamefont {Dap}}, \bibinfo {author} {\bibfnamefont {R.}~\bibnamefont {Clergereaux}},\ and\ \bibinfo {author} {\bibfnamefont {O.}~\bibnamefont {Pascal}},\ }\bibfield  {title} {\bibinfo {title} {Plasma generation using time reversal of microwaves},\ }\href@noop {} {\bibfield  {journal} {\bibinfo  {journal} {Applied Physics Letters}\ }\textbf {\bibinfo {volume} {115}} (\bibinfo {year} {2019})}\BibitemShut {NoStop}%
\bibitem [{\citenamefont {Mazieres}\ \emph {et~al.}(2020)\citenamefont {Mazieres}, \citenamefont {Pascaud}, \citenamefont {Pascal}, \citenamefont {Clergereaux}, \citenamefont {Stafford}, \citenamefont {Dap},\ and\ \citenamefont {Liard}}]{mazieres2020spatio}%
  \BibitemOpen
  \bibfield  {author} {\bibinfo {author} {\bibfnamefont {V.}~\bibnamefont {Mazieres}}, \bibinfo {author} {\bibfnamefont {R.}~\bibnamefont {Pascaud}}, \bibinfo {author} {\bibfnamefont {O.}~\bibnamefont {Pascal}}, \bibinfo {author} {\bibfnamefont {R.}~\bibnamefont {Clergereaux}}, \bibinfo {author} {\bibfnamefont {L.}~\bibnamefont {Stafford}}, \bibinfo {author} {\bibfnamefont {S.}~\bibnamefont {Dap}},\ and\ \bibinfo {author} {\bibfnamefont {L.}~\bibnamefont {Liard}},\ }\bibfield  {title} {\bibinfo {title} {Spatio-temporal dynamics of a nanosecond pulsed microwave plasma ignited by time reversal},\ }\href@noop {} {\bibfield  {journal} {\bibinfo  {journal} {Plasma Sources Science and Technology}\ }\textbf {\bibinfo {volume} {29}},\ \bibinfo {pages} {125017} (\bibinfo {year} {2020})}\BibitemShut {NoStop}%
\bibitem [{\citenamefont {Mazi{\`e}res}\ \emph {et~al.}(2021)\citenamefont {Mazi{\`e}res}, \citenamefont {Pascal}, \citenamefont {Pascaud}, \citenamefont {Liard}, \citenamefont {Dap}, \citenamefont {Clergereaux},\ and\ \citenamefont {Boeuf}}]{mazieres2021space}%
  \BibitemOpen
  \bibfield  {author} {\bibinfo {author} {\bibfnamefont {V.}~\bibnamefont {Mazi{\`e}res}}, \bibinfo {author} {\bibfnamefont {O.}~\bibnamefont {Pascal}}, \bibinfo {author} {\bibfnamefont {R.}~\bibnamefont {Pascaud}}, \bibinfo {author} {\bibfnamefont {L.}~\bibnamefont {Liard}}, \bibinfo {author} {\bibfnamefont {S.}~\bibnamefont {Dap}}, \bibinfo {author} {\bibfnamefont {R.}~\bibnamefont {Clergereaux}},\ and\ \bibinfo {author} {\bibfnamefont {J.-P.}\ \bibnamefont {Boeuf}},\ }\bibfield  {title} {\bibinfo {title} {Space-time plasma-steering source: Control of microwave plasmas in overmoded cavities},\ }\href@noop {} {\bibfield  {journal} {\bibinfo  {journal} {Physical Review Applied}\ }\textbf {\bibinfo {volume} {16}},\ \bibinfo {pages} {054038} (\bibinfo {year} {2021})}\BibitemShut {NoStop}%
\bibitem [{\citenamefont {Abramowitz}\ and\ \citenamefont {Stegun}(1948)}]{abramowitz1948handbook}%
  \BibitemOpen
  \bibfield  {author} {\bibinfo {author} {\bibfnamefont {M.}~\bibnamefont {Abramowitz}}\ and\ \bibinfo {author} {\bibfnamefont {I.~A.}\ \bibnamefont {Stegun}},\ }\href@noop {} {\emph {\bibinfo {title} {Handbook of mathematical functions with formulas, graphs, and mathematical tables}}},\ Vol.~\bibinfo {volume} {55}\ (\bibinfo  {publisher} {US Government printing office},\ \bibinfo {year} {1948})\BibitemShut {NoStop}%
\bibitem [{\citenamefont {Kourtzanidis}\ \emph {et~al.}(2015)\citenamefont {Kourtzanidis}, \citenamefont {Rogier},\ and\ \citenamefont {Boeuf}}]{kourtzanidis2015adi}%
  \BibitemOpen
  \bibfield  {author} {\bibinfo {author} {\bibfnamefont {K.}~\bibnamefont {Kourtzanidis}}, \bibinfo {author} {\bibfnamefont {F.}~\bibnamefont {Rogier}},\ and\ \bibinfo {author} {\bibfnamefont {J.-P.}\ \bibnamefont {Boeuf}},\ }\bibfield  {title} {\bibinfo {title} {Adi-fdtd modeling of microwave plasma discharges in air towards fully three-dimensional simulations},\ }\href@noop {} {\bibfield  {journal} {\bibinfo  {journal} {Computer Physics Communications}\ }\textbf {\bibinfo {volume} {195}},\ \bibinfo {pages} {49} (\bibinfo {year} {2015})}\BibitemShut {NoStop}%
\bibitem [{\citenamefont {MacDonald}(1966)}]{macdonald1966microwave}%
  \BibitemOpen
  \bibfield  {author} {\bibinfo {author} {\bibfnamefont {A.~D.}\ \bibnamefont {MacDonald}},\ }\href@noop {} {\emph {\bibinfo {title} {Microwave breakdown in gases}}},\ Wiley series in plasma physics\ (\bibinfo  {publisher} {Wiley},\ \bibinfo {address} {New York},\ \bibinfo {year} {1966})\BibitemShut {NoStop}%
\bibitem [{\citenamefont {Taflove}\ and\ \citenamefont {Hagness}(2000)}]{taflove2000computational}%
  \BibitemOpen
  \bibfield  {author} {\bibinfo {author} {\bibfnamefont {A.}~\bibnamefont {Taflove}}\ and\ \bibinfo {author} {\bibfnamefont {S.~C.}\ \bibnamefont {Hagness}},\ }\bibfield  {title} {\bibinfo {title} {Computational electrodynamics},\ }\href@noop {} {\bibfield  {journal} {\bibinfo  {journal} {The Finite-Difference Time-Domain Method}\ } (\bibinfo {year} {2000})}\BibitemShut {NoStop}%
\bibitem [{\citenamefont {Belkhir}(2008)}]{belkhir2008extension}%
  \BibitemOpen
  \bibfield  {author} {\bibinfo {author} {\bibfnamefont {A.}~\bibnamefont {Belkhir}},\ }\emph {\bibinfo {title} {Extension de la mod{\'e}lisation par FDTD en nano-optique}},\ \href@noop {} {Ph.D. thesis},\ \bibinfo  {school} {Universit{\'e} de Franche-Comt{\'e}} (\bibinfo {year} {2008})\BibitemShut {NoStop}%
\bibitem [{\citenamefont {Hadi}\ \emph {et~al.}(2016)\citenamefont {Hadi}, \citenamefont {Elsherbeni}, \citenamefont {Piket-May},\ and\ \citenamefont {Mahmoud}}]{hadi2016radial}%
  \BibitemOpen
  \bibfield  {author} {\bibinfo {author} {\bibfnamefont {M.~F.}\ \bibnamefont {Hadi}}, \bibinfo {author} {\bibfnamefont {A.~Z.}\ \bibnamefont {Elsherbeni}}, \bibinfo {author} {\bibfnamefont {M.~J.}\ \bibnamefont {Piket-May}},\ and\ \bibinfo {author} {\bibfnamefont {S.~F.}\ \bibnamefont {Mahmoud}},\ }\bibfield  {title} {\bibinfo {title} {Radial waves based dispersion analysis of the body-of-revolution fdtd method},\ }\href@noop {} {\bibfield  {journal} {\bibinfo  {journal} {IEEE Transactions on Antennas and Propagation}\ }\textbf {\bibinfo {volume} {65}},\ \bibinfo {pages} {721} (\bibinfo {year} {2016})}\BibitemShut {NoStop}%
\bibitem [{\citenamefont {Tierens}\ and\ \citenamefont {De~Zutter}(2011)}]{tierens2011time}%
  \BibitemOpen
  \bibfield  {author} {\bibinfo {author} {\bibfnamefont {W.}~\bibnamefont {Tierens}}\ and\ \bibinfo {author} {\bibfnamefont {D.}~\bibnamefont {De~Zutter}},\ }\bibfield  {title} {\bibinfo {title} {Time-domain formulation of cold plasma based on mass-lumped finite elements},\ }in\ \href@noop {} {\emph {\bibinfo {booktitle} {AIP Conference Proceedings}}},\ Vol.\ \bibinfo {volume} {1406}\ (\bibinfo {organization} {American Institute of Physics},\ \bibinfo {year} {2011})\ pp.\ \bibinfo {pages} {381--384}\BibitemShut {NoStop}%
\bibitem [{\citenamefont {Wu}\ \emph {et~al.}(2024)\citenamefont {Wu}, \citenamefont {Liu},\ and\ \citenamefont {Zhou}}]{wu2024vortex}%
  \BibitemOpen
  \bibfield  {author} {\bibinfo {author} {\bibfnamefont {Y.}~\bibnamefont {Wu}}, \bibinfo {author} {\bibfnamefont {M.}~\bibnamefont {Liu}},\ and\ \bibinfo {author} {\bibfnamefont {C.}~\bibnamefont {Zhou}},\ }\bibfield  {title} {\bibinfo {title} {Vortex electromagnetic wave propagation in magnetized plasma},\ }\href@noop {} {\bibfield  {journal} {\bibinfo  {journal} {IEEE Transactions on Antennas and Propagation}\ } (\bibinfo {year} {2024})}\BibitemShut {NoStop}%
\bibitem [{\citenamefont {Delage}\ \emph {et~al.}(2022)\citenamefont {Delage}, \citenamefont {Pascal}, \citenamefont {Sokoloff},\ and\ \citenamefont {Mazi{\`e}res}}]{delage2022experimental}%
  \BibitemOpen
  \bibfield  {author} {\bibinfo {author} {\bibfnamefont {T.}~\bibnamefont {Delage}}, \bibinfo {author} {\bibfnamefont {O.}~\bibnamefont {Pascal}}, \bibinfo {author} {\bibfnamefont {J.}~\bibnamefont {Sokoloff}},\ and\ \bibinfo {author} {\bibfnamefont {V.}~\bibnamefont {Mazi{\`e}res}},\ }\bibfield  {title} {\bibinfo {title} {Experimental demonstration of virtual critical coupling to a single-mode microwave cavity},\ }\href@noop {} {\bibfield  {journal} {\bibinfo  {journal} {Journal of Applied Physics}\ }\textbf {\bibinfo {volume} {132}} (\bibinfo {year} {2022})}\BibitemShut {NoStop}%
\bibitem [{\citenamefont {Kim}\ \emph {et~al.}(2016)\citenamefont {Kim}, \citenamefont {Badsha}, \citenamefont {Yoon}, \citenamefont {Lee}, \citenamefont {Jun},\ and\ \citenamefont {Hwangbo}}]{kim2016general}%
  \BibitemOpen
  \bibfield  {author} {\bibinfo {author} {\bibfnamefont {T.~Y.}\ \bibnamefont {Kim}}, \bibinfo {author} {\bibfnamefont {M.~A.}\ \bibnamefont {Badsha}}, \bibinfo {author} {\bibfnamefont {J.}~\bibnamefont {Yoon}}, \bibinfo {author} {\bibfnamefont {S.~Y.}\ \bibnamefont {Lee}}, \bibinfo {author} {\bibfnamefont {Y.~C.}\ \bibnamefont {Jun}},\ and\ \bibinfo {author} {\bibfnamefont {C.~K.}\ \bibnamefont {Hwangbo}},\ }\bibfield  {title} {\bibinfo {title} {General strategy for broadband coherent perfect absorption and multi-wavelength all-optical switching based on epsilon-near-zero multilayer films},\ }\href@noop {} {\bibfield  {journal} {\bibinfo  {journal} {Scientific Reports}\ }\textbf {\bibinfo {volume} {6}},\ \bibinfo {pages} {22941} (\bibinfo {year} {2016})}\BibitemShut {NoStop}%
\bibitem [{\citenamefont {Islam}\ \emph {et~al.}(2023)\citenamefont {Islam}, \citenamefont {Gaire}, \citenamefont {Madanayake},\ and\ \citenamefont {Bhardwaj}}]{islam2023generation}%
  \BibitemOpen
  \bibfield  {author} {\bibinfo {author} {\bibfnamefont {M.~K.}\ \bibnamefont {Islam}}, \bibinfo {author} {\bibfnamefont {P.}~\bibnamefont {Gaire}}, \bibinfo {author} {\bibfnamefont {A.}~\bibnamefont {Madanayake}},\ and\ \bibinfo {author} {\bibfnamefont {S.}~\bibnamefont {Bhardwaj}},\ }\bibfield  {title} {\bibinfo {title} {Generation of vector vortex wave modes in cylindrical waveguides},\ }\href@noop {} {\bibfield  {journal} {\bibinfo  {journal} {Scientific Reports}\ }\textbf {\bibinfo {volume} {13}},\ \bibinfo {pages} {11066} (\bibinfo {year} {2023})}\BibitemShut {NoStop}%
\bibitem [{\citenamefont {Hirsh}(2012)}]{hirsh2012gaseous}%
  \BibitemOpen
  \bibfield  {author} {\bibinfo {author} {\bibfnamefont {M.}~\bibnamefont {Hirsh}},\ }\href@noop {} {\emph {\bibinfo {title} {Gaseous electronics}}},\ Vol.~\bibinfo {volume} {1}\ (\bibinfo  {publisher} {Elsevier},\ \bibinfo {year} {2012})\BibitemShut {NoStop}%
\bibitem [{\citenamefont {Fu}\ \emph {et~al.}(2017)\citenamefont {Fu}, \citenamefont {Wang}, \citenamefont {Zou}, \citenamefont {Yang}, \citenamefont {Verboncoeur},\ and\ \citenamefont {Christlieb}}]{fu2017investigation}%
  \BibitemOpen
  \bibfield  {author} {\bibinfo {author} {\bibfnamefont {Y.}~\bibnamefont {Fu}}, \bibinfo {author} {\bibfnamefont {X.}~\bibnamefont {Wang}}, \bibinfo {author} {\bibfnamefont {X.}~\bibnamefont {Zou}}, \bibinfo {author} {\bibfnamefont {S.}~\bibnamefont {Yang}}, \bibinfo {author} {\bibfnamefont {J.~P.}\ \bibnamefont {Verboncoeur}},\ and\ \bibinfo {author} {\bibfnamefont {A.~J.}\ \bibnamefont {Christlieb}},\ }\bibfield  {title} {\bibinfo {title} {Investigation on the similarity law of low-pressure glow discharges based on the light intensity distributions in geometrically similar gaps},\ }\href@noop {} {\bibfield  {journal} {\bibinfo  {journal} {Physics of Plasmas}\ }\textbf {\bibinfo {volume} {24}} (\bibinfo {year} {2017})}\BibitemShut {NoStop}%
\bibitem [{\citenamefont {Pugliese}\ \emph {et~al.}(2015)\citenamefont {Pugliese}, \citenamefont {Meucci}, \citenamefont {Euzzor}, \citenamefont {Freire},\ and\ \citenamefont {Gallas}}]{pugliese2015complex}%
  \BibitemOpen
  \bibfield  {author} {\bibinfo {author} {\bibfnamefont {E.}~\bibnamefont {Pugliese}}, \bibinfo {author} {\bibfnamefont {R.}~\bibnamefont {Meucci}}, \bibinfo {author} {\bibfnamefont {S.}~\bibnamefont {Euzzor}}, \bibinfo {author} {\bibfnamefont {J.~G.}\ \bibnamefont {Freire}},\ and\ \bibinfo {author} {\bibfnamefont {J.~A.}\ \bibnamefont {Gallas}},\ }\bibfield  {title} {\bibinfo {title} {Complex dynamics of a dc glow discharge tube: Experimental modeling and stability diagrams},\ }\href@noop {} {\bibfield  {journal} {\bibinfo  {journal} {Scientific Reports}\ }\textbf {\bibinfo {volume} {5}},\ \bibinfo {pages} {8447} (\bibinfo {year} {2015})}\BibitemShut {NoStop}%
\bibitem [{\citenamefont {Chiad}\ \emph {et~al.}(2009)\citenamefont {Chiad}, \citenamefont {Al-Zubaydi}, \citenamefont {Khalaf},\ and\ \citenamefont {Khudiar}}]{chiad2009construction}%
  \BibitemOpen
  \bibfield  {author} {\bibinfo {author} {\bibfnamefont {B.~T.}\ \bibnamefont {Chiad}}, \bibinfo {author} {\bibfnamefont {T.}~\bibnamefont {Al-Zubaydi}}, \bibinfo {author} {\bibfnamefont {M.}~\bibnamefont {Khalaf}},\ and\ \bibinfo {author} {\bibfnamefont {A.}~\bibnamefont {Khudiar}},\ }\bibfield  {title} {\bibinfo {title} {Construction and characterization of a low pressure plasma reactor using dc glow discharge},\ }\href@noop {} {\bibfield  {journal} {\bibinfo  {journal} {Journal of Optoelectronics and Biomedical Materials}\ }\textbf {\bibinfo {volume} {1}},\ \bibinfo {pages} {255} (\bibinfo {year} {2009})}\BibitemShut {NoStop}%
\bibitem [{\citenamefont {Fleury}\ \emph {et~al.}(2015)\citenamefont {Fleury}, \citenamefont {Monticone},\ and\ \citenamefont {Al{\`u}}}]{fleury2015invisibility}%
  \BibitemOpen
  \bibfield  {author} {\bibinfo {author} {\bibfnamefont {R.}~\bibnamefont {Fleury}}, \bibinfo {author} {\bibfnamefont {F.}~\bibnamefont {Monticone}},\ and\ \bibinfo {author} {\bibfnamefont {A.}~\bibnamefont {Al{\`u}}},\ }\bibfield  {title} {\bibinfo {title} {Invisibility and cloaking: Origins, present, and future perspectives},\ }\href@noop {} {\bibfield  {journal} {\bibinfo  {journal} {Physical Review Applied}\ }\textbf {\bibinfo {volume} {4}},\ \bibinfo {pages} {037001} (\bibinfo {year} {2015})}\BibitemShut {NoStop}%
\bibitem [{\citenamefont {Sakai}\ \emph {et~al.}(2016)\citenamefont {Sakai}, \citenamefont {Yamaguchi}, \citenamefont {Bambina}, \citenamefont {Iwai}, \citenamefont {Nakamura}, \citenamefont {Tamayama},\ and\ \citenamefont {Miyagi}}]{sakai2016plasma}%
  \BibitemOpen
  \bibfield  {author} {\bibinfo {author} {\bibfnamefont {O.}~\bibnamefont {Sakai}}, \bibinfo {author} {\bibfnamefont {S.}~\bibnamefont {Yamaguchi}}, \bibinfo {author} {\bibfnamefont {A.}~\bibnamefont {Bambina}}, \bibinfo {author} {\bibfnamefont {A.}~\bibnamefont {Iwai}}, \bibinfo {author} {\bibfnamefont {Y.}~\bibnamefont {Nakamura}}, \bibinfo {author} {\bibfnamefont {Y.}~\bibnamefont {Tamayama}},\ and\ \bibinfo {author} {\bibfnamefont {S.}~\bibnamefont {Miyagi}},\ }\bibfield  {title} {\bibinfo {title} {Plasma metamaterials as cloaking and nonlinear media},\ }\href@noop {} {\bibfield  {journal} {\bibinfo  {journal} {Plasma Physics and Controlled Fusion}\ }\textbf {\bibinfo {volume} {59}},\ \bibinfo {pages} {014042} (\bibinfo {year} {2016})}\BibitemShut {NoStop}%
\bibitem [{\citenamefont {Chaudhury}\ and\ \citenamefont {Boeuf}(2010)}]{chaudhury2010computational}%
  \BibitemOpen
  \bibfield  {author} {\bibinfo {author} {\bibfnamefont {B.}~\bibnamefont {Chaudhury}}\ and\ \bibinfo {author} {\bibfnamefont {J.-P.}\ \bibnamefont {Boeuf}},\ }\bibfield  {title} {\bibinfo {title} {Computational studies of filamentary pattern formation in a high power microwave breakdown generated air plasma},\ }\href@noop {} {\bibfield  {journal} {\bibinfo  {journal} {IEEE transactions on plasma science}\ }\textbf {\bibinfo {volume} {38}},\ \bibinfo {pages} {2281} (\bibinfo {year} {2010})}\BibitemShut {NoStop}%
\end{thebibliography}%

\appendix

\section{Discretized equations of the numerical model}\label{ap:disc}

\begin{figure}[h]
    \includegraphics[width=8cm]{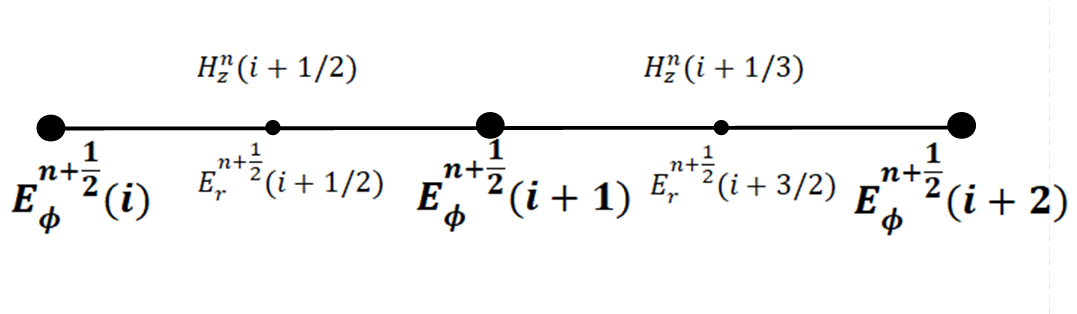}
    \caption{Yee grid of the field \textbf{E} and \textbf{H} in the case of 1D.}
    \label{fig:Yee}
\end{figure}

The different components of the electron velocity \(\mathbf{v_e}\) and \textbf{e} fields are defined at the same position in the spatial grid \cite{mazieres2021space,chaudhury2010computational}. However, the \(h_z\) and \(e_r\) fields must also be defined at the same positions. Indeed, from the equations (\ref{ErBOR}) and (\ref{HzBOR}) show it is possible to show that they must be set to the same positions and that \(e_\phi\) must be set to a different location than \(h_z\) and \(e_r\) \footnote{The boundary conditions used in this study for \(m=1\) (Here, \(m=1\) is chosen because there are no solutions for \(s=-1\) or \(s=0\) when \(m=0\)) at \(r=0\) are: \(H_z^{n}(1/2) = 0\), \(e_\phi^{n+1/2}(1/2) = 0\), and \(e_r^{n+1/2}(0) = e_r^{n+1/2}(1/2)\).}. The Yee grid used in this study is represented in Fig. \ref{fig:Yee}. However, from a temporal point of view, the \textbf{e} fields are updated at the \(n+1/2\) time step, while the \textbf{h} fields are updated at the \(n\) time step.

Due to the use of finite centered differences for the derivatives in numerical simulations, numerical instability can occur because of the discretization of these continuous equations. This can lead to a phenomenon where the wave travels multiple cells between two time intervals \(\Delta t\). A stability criterion is then use to negate this effect, where the spatial step \(\Delta d\) and the time step \(\Delta t\) must be such that the wave cannot travel more than one cell per time step. Thus, the numerical stability criterion is given in the case of BOR-FDTD by the CFL criterion \(\Delta t = \frac{\Delta d}{c(m+1)}\), with \(c\) being the speed of light \cite{taflove2000computational}.

Equations (\ref{M_F}) to (\ref{M_A}) yield equations (\ref{Er}-\ref{Hz}) in a cylindrical case where the cylinder is considered semi-infinite.

\begin{align}
\begin{split}
    &\varepsilon_0\frac{\partial E_r}{\partial t} = \frac{1}{r}\frac{\partial H_z}{\partial \phi} + J_{pr} 
    \label{Er}
\end{split} \\
\begin{split}
    & \varepsilon_0\frac{\partial E_\phi}{\partial t} = - \frac{\partial H_z}{\partial r} + J_{p\phi} 
    \label{Ephi}
\end{split} \\
\begin{split}
     & \mu_0\frac{\partial H_z}{\partial t} = -\frac{1}{r}\left(\frac{\partial (rE_\phi)}{\partial r} - \frac{\partial E_r}{\partial \phi}\right)
     \label{Hz}
\end{split}
\end{align}

Thanks to the symmetry of revolution, it is possible to decompose the electric and magnetic fields into a Fourier series \cite{taflove2000computational}:

\begin{align}
    \mathbf{E}(r,\phi,t) =  \sum^n \mathbf{e_{\mu}}(r,t) \cos(m\phi) +  \mathbf{e_{\nu}}(r,t) \sin(m\phi)   \label{FourierE} \\
    \mathbf{H}(r,\phi,t) =   \sum^n \mathbf{h_\mu}(r,t) \cos(m\phi) +  \mathbf{h_\nu}(r,t) \sin(m\phi)   \label{FourierH}
\end{align}

With \(\mathbf{e_{\mu/\nu}}\) (resp. \(\mathbf{h_{\mu/\nu}}\)), the electric (resp. magnetic) field without the \(\phi\) dependence. In our case it is considered that the wave has only one azimuthal mode. Thus, injecting the equations (\ref{FourierE}) and (\ref{FourierH}) into equations (\ref{Er}) to (\ref{Hz}) and following the development given by \cite{taflove2000computational} leads to (for simplification, the notation of the subtract \(\mu\) and \(\nu\) has been taken out) :

\begin{align}
    \label{ErBOR}
    &\varepsilon_0\frac{\partial e_r}{\partial t} = -\frac{m}{r}h_z -en_e v_{er}  \\
    & \varepsilon_0\frac{\partial e_\phi}{\partial t} =  - \frac{\partial h_z}{\partial r} -en_e v_{e\phi} \label{EphiBor}\\
     &-\mu_0\frac{\partial h_z}{\partial t} = \frac{1}{r}\left(\frac{\partial (re_\phi)}{\partial r} -  me_r\right)  
     \label{HzBOR}
\end{align}

The same method is used to discretize the equations (\ref{ErBOR}) to (\ref{HzBOR}) as what can be found for a classical FDTD simulation \cite{chaudhury2010computational}.

\begin{align}
\begin{split}
\label{Hz_disc}
h_z^{n+1}(i+1/2) &= h_z^{n}(i+1/2) \\ &\quad + \frac{-\Delta t}{\mu_0 \Delta d(i+1/2)} \\ &\quad \left[ (i+1) e_\phi^{n+1/2}(i+1) 
- i e_\phi^{n+1/2}(i) \right] \\
&\quad - \frac{m \Delta t}{\mu_0 \Delta d(i+1/2)} e_r^{n+1/2}(i+1/2)
\end{split} \\
\begin{split}
\label{Er_disc}
e_r^{n+1/2}(i + 1/2) &= \frac{1- \beta}{1 + \beta} e_r^{n-1/2}(i + 1/2) \\ &\quad + \frac{en_e \Delta t}{2 \varepsilon_0} \frac{1 + \alpha}{1 + \beta} v_{er}^{n-1/2}(i + 1/2) \\
&\quad + \frac{m}{\Delta d(i + 1/2)} \frac{\Delta t}{\varepsilon_0 (1 + \beta)} h_z^{n}(i + 1/2)
\end{split} \\
\begin{split}
\label{Ephi_disc}
e_\phi^{n+1/2}(i+1) &= \frac{1 - \beta}{1 + \beta} e_\phi^{n+1/2}(i) + \frac{en_e \Delta t}{2 \varepsilon_0} \frac{1 + \alpha}{1 + \beta} v_{e\phi}^{n-1/2} \\
&\quad - \frac{\Delta t}{\varepsilon_0 \Delta d (1 + \beta)} \left( h_z^{n}(i+1) - h_z^{n}(i) \right)
\end{split} \\
\begin{split}
\label{ve_disc}
\text{For } \ell = r,\phi,\quad v_{e,\ell}^{n+1/2} &= \alpha v_{e,\ell}^{n-1/2} - \frac{e \Delta t}{2 m_e \zeta} \left( e_{\ell}^{n+1} + e_{\ell}^n \right)
\end{split}
\end{align}
with :
\begin{align}
\begin{split} 
\alpha=\frac{1-a}{1+a},~\zeta=1+a,~a=\frac{\nu_m\Delta_t}{2}\\
\text{and}~\beta=\frac{\omega_p^2\Delta_t^2}{4\zeta}
\end{split}
\end{align}

Equations (\ref{Hz_disc}) to (\ref{ve_disc}) correspond to the discretized versions of the equations (\ref{ve}) and (\ref{ErBOR}) to (\ref{HzBOR}). The algorithm will transcribe the spatio-temporal evolution of the \textbf{e} and \textbf{h} fields and the electron velocity \(\mathbf{v_e}\)

\end{document}